%% file: sblind_vb_57_submit_figDir_arXiv.tex
\documentclass[review,10pt]{elsarticle}

\usepackage{dsfont}
\usepackage{amsmath}
\usepackage{amssymb}
\usepackage{amsfonts}
\usepackage{graphicx}
\usepackage{amsmath}
\usepackage{multirow}
\usepackage{comment}
\usepackage{algorithm}
\usepackage{algorithmic}
\usepackage{threeparttable}

\usepackage[noadjust]{cite}

\usepackage{color}
\usepackage{array}
\usepackage{color,amssymb,amsmath,array,subfigure,amsthm}

\usepackage{setspace}
\usepackage[noadjust]{cite}

\begin{document}

    \input notations.tex
    \newcommand{\bydef}{\stackrel{{def}}{=}}
    \newcommand{\ici}{$\blacktriangleright \;$}
    \newcommand{\esp}{\mathsf{E}}
    \newcommand{\doigt}{\noindent \Pisymbol{pzd}{43}}
    \newcommand{\indicfun}[2]{\mathbf{1}_{#1}\left(#2\right)}
    \newcommand{\Proba}{\mathrm{P}}

\newcommand{\figextension}{pdf}
\newcommand{\figwidth}{0.85\columnwidth}
\newcommand{\figwidthbis}{0.99\columnwidth}
\newcommand{\figwidthsmall}{0.35\columnwidth}
\newcommand{\figwidthFigThree}{0.25\columnwidth}
\newcommand{\figwidthsmalldouble}{0.4675\columnwidth}
\newcommand{\figwidthbig}{\figwidthsmall}
\newcommand{\red}[1]{\textcolor[rgb]{0.00,0.00,0.00}{#1}}

\def\EE{{\mathrm{E}}}


\begin{frontmatter}

\title{Variational Semi-blind Sparse Deconvolution with Orthogonal Kernel Bases and its Application to MRFM}

\author[umich]{Se Un Park\corref{cor1}}
\ead{seunpark@umich.edu}
\author[toul]{Nicolas Dobigeon}
\ead{nicolas.dobigeon@enseeiht.fr}
\author[umich]{Alfred O. Hero}
\ead{hero@umich.edu}

\cortext[cor1]{Corresponding author. Tel: +1 (734) 763-4497, fax: +1 (734) 763-8041  \\
{This research was partially supported by a grant from ARO, grant number W911NF-05-1-0403.}
}

\address[umich]{University of Michigan, Department of EECS, Ann
Arbor, MI 48109-2122, USA    }
\address[toul]{University of Toulouse, IRIT/INP-ENSEEIHT, 2 rue
Camichel, BP 7122, 31071 Toulouse cedex 7, France}





\begin{abstract}
We present a variational Bayesian method of joint image reconstruction and point
spread function (PSF) estimation when the PSF of the imaging device
is only partially known. To solve this semi-blind
deconvolution problem, prior distributions are specified for the PSF and the 3D image.
Joint image reconstruction and PSF estimation is then performed within a
Bayesian framework, using a variational algorithm to estimate the
posterior distribution.
The image prior distribution imposes an explicit atomic measure that corresponds to image sparsity.
Importantly, the proposed Bayesian deconvolution algorithm does not require hand tuning. Simulation results clearly demonstrate that the
semi-blind deconvolution algorithm compares favorably with
previous Markov chain Monte Carlo (MCMC) version of myopic sparse
reconstruction. \red{It significantly outperforms mismatched non-blind algorithms that rely
on the assumption of the perfect knowledge of the PSF.}
The algorithm is illustrated on real data from magnetic resonance force microscopy (MRFM).
\end{abstract}

\begin{keyword}
Variational Bayesian inference, posterior image distribution, image reconstruction, hyperparameter estimation, MRFM experiment.
\end{keyword}

\end{frontmatter}

\section{Introduction}

The standard and popular image deconvolution techniques generally
assume that the space-invariant instrument response, i.e., the point spread function
(PSF), is perfectly known. However, in many practical situations,
the true PSF is either unknown or, at best, partially known. 
\red{ For example, in an optical system a perfectly known PSF does not exist because of light diffraction, apparatus/lense aberration, out-of-focus, or image motion \citep{Ward1987,Kundur1996}. Such imperfections are common in general imaging systems including MRFM, where there exist additional model PSF errors in the sensitive magnetic resonance condition \citep{Mamin2003}.
}
In such circumstances, the PSF required in the reconstruction process
is mismatched with the true PSF. The quality of standard image
reconstruction techniques may suffer from this disparity. To deal
with this mismatch, deconvolution methods have been proposed to
estimate the unknown image and the PSF jointly. When prior
knowledge of the PSF is available, these methods are usually
referred to as semi-blind deconvolution \citep{Makni2005, Pillonetto2007} or myopic
deconvolution \citep{Sarri1998, Chenegros2007, Park2011}.

In this paper, we formulate the semi-blind deconvolution task as an
estimation problem in a Bayesian setting.
Bayesian estimation
has the great advantage of offering a flexible framework to solve
complex model-based problems. Prior
information available on the parameters to be estimated can be
efficiently included within the model, leading to an implicit
regularization of our ill-posed problem. In addition, the Bayes framework produces posterior estimates of uncertainty, via posterior variance and posterior confidence intervals.
Extending our previous work, %
we propose a variational estimator for the parameters
 as contrasted to the Monte Carlo approach in \citep{Park2012}. This extension is non-trivial.  Our variational Bayes algorithm iterates on a hidden variable domain associated with the mixture coefficients. \red{ This algorithm is} faster, more scalable for equivalent image reconstruction qualities in \citep{Park2012}.

Like in \citep{Park2012}, the PSF uncertainty is modeled as the deviation of the a priori known PSF
from the true PSF.
Applying an eigendecomposition to the PSF covariance,
the deviation is represented as a linear
combination of orthogonal PSF bases with unknown coefficients that need to be
estimated.
Furthermore, we assume the desired image is sparse,
corresponding to the natural sparsity of the molecular
image. The image prior is a weighted sum of a sparsity
inducing part and a continuous distribution; a positive truncated \emph{Laplacian and atom at zero} (LAZE)
prior\footnote{\red{A Laplace distribution as a prior distribution acts as a sparse regularization using $\ell_1$ norm. This can be seen by taking negative logarithm on the distribution.}} \citep{Ting2009}. Similar priors have been applied to estimating mixtures of densities
 \citep{Bishop2006, Nasios2006, Corduneanu2001} \red{and sparse, nonnegative hyperspectral unmixing \citep{Themelis2012}}. Here we introduce a hidden label variable for the contribution of the discrete mass (empty pixel) and a
continuous density function (non-empty pixel).
\red{Similar to our `hybrid' mixture model, inhomogeneous gamma-Gaussian mixture models have been proposed in \citep{Makni2008}.}

Bayesian inference of parameters from the posterior distribution generally
requires challenging computations, such as functional optimization and
numerical integration.
One widely advocated strategy relies on approximations to the
minimum mean square error (MMSE)
or maximum a posteriori (MAP) estimators using samples drawn from
the posterior distribution. Generation of these samples can be
accomplished using Markov chain Monte Carlo  methods (MCMC)
 \citep{Robert2004}. MCMC has been successfully adopted in
numerous imaging problems such as image segmentation, denoising,
and deblurring \citep{Gilks1999,Robert2004}. %
Recently, to solve blind deconvolution, two promising
semi-blind MCMC methods have been
suggested
 \citep{Park2012,Orieux2010}. However, these sampling methods  have the disadvantage that convergence may be slow.

An alternative to Monte Carlo integration is a
variational approximation to the posterior distribution, and this approach is adopted in this paper. These
approximations have been extensively exploited to conduct inference in graphical
models \citep{Attias00}. If properly designed, they can produce an
analytical posterior distribution from which Bayesian estimators
can be efficiently computed. Compared to MCMC, variational
methods are 
of lower computational complexity,
since they avoid stochastic simulation.
However, variational Bayes (VB) approaches have intrinsic limits; the convergence to the true distribution is not guaranteed, \red{even though the posterior distribution will be asymptotically normal with mean equal to the maximum likelihood estimator under suitable conditions \citep{Walker1969}. 
In addition, variational Bayes approximations can be easily implemented for only a limited number of statistical models.}
For example, this method is difficult to apply when latent variables have distributions that do not belong to the exponential family \red{(e.g. a discrete distribution \citep{Park2012}).
For mixture distributions,
variational estimators in Gaussian mixtures and in exponential family converge locally to maximum likelihood estimator  \citep{Wang2004a,Wang2006}.
The theoretical convergence properties for sparse mixture models, such as our proposed model, are as yet unknown. This has not hindered the application of VB to sparse models to problems in  our sparse image mixture model.}
\red{
Another possible intrinsic limit of the variational Bayes approach, particularly in (semi)-blind deconvolution, is that the posterior covariance structure cannot  be effectively estimated nor recovered, unless the true joint distributions have independent individual distributions. This is primarily because VB algorithms are based on minimizing the KL-divergence between
 the true distribution and the VB approximating distribution, which is assumed to be factorized with respect to the individual parameters.
}

However, despite these limits, VB approaches have been widely applied with success to many different engineering problems \citep{Bishop2000,Ghahramani2000,Beal2003,Winn2005}.
A principal contribution of this paper is the development and implementation of a VB algorithm for mixture distributions in a hierarchical Bayesian model.
Similarly, the framework permits a Gaussian prior \citep{Babacan2009} or a Student's-t prior   \citep{Tzikas2009} for the PSF.
We present comparisons of our variational solution to other blind deconvolution methods.
These include the total variation (TV) prior for the PSF \citep{Amizic2010} and
natural sharp edge priors for images with PSF regularization \citep{Almeida2010}. We also compare to basis kernels \citep{Tzikas2009}, the mixture model algorithm of Fergus \emph{et al.} \citep{Fergus2006},
and the related method of Shan \emph{et al.} \citep{Shan2008} under a motion blur model.



To implement variational Bayesian inference, prior
distributions and the instrument-dependent likelihood function are specified.
Then the posterior distributions are estimated by minimizing the Kullback-Leibler (KL) distance between the model and the empirical distribution.
Simulations conducted on synthetic images show that the resulting
myopic deconvolution algorithm outperforms previous mismatched non-blind
algorithms and competes with the previous MCMC-based semi-blind
method \citep{Park2012} with lower computational complexity.

We illustrate the proposed method on real data from magnetic resonance
force microscopy (MRFM) experiments.
MRFM is an emerging molecular imaging modality that has the potential for achieving $3$D
atomic scale resolution \citep{Sidles1991, Sidles1992, Sidles1995}.
Recently, MRFM has successfully  demonstrated imaging
 \citep{Rugar1992,Zuger1996} of a tobacco mosaic virus \citep{Degen2009}.
The $3$D image reconstruction problem for MRFM experiments was
investigated with Wiener filters
 \citep{Zuger1993,Zuger1994,Zuger1996}, iterative least square
reconstruction approaches \citep{Chao2004,Degen2009,Degen2009compl},
and recently the Bayesian estimation framework
 \citep{Ting2009,Dobigeon2009a,Park2011,Park2012}. The drawback of these approaches is that they
require prior
knowledge \red{on} the PSF. However, in many
practical situations of MRFM imaging, the exact PSF, i.e., the
response of the MRFM tip, is only partially known \citep{Mamin2003}.
The proposed semi-blind reconstruction method accounts for this partial knowledge.

The rest of this paper is organized as follows. Section
\ref{sec:formulation} formulates the imaging deconvolution problem
in a hierarchical Bayesian framework. Section \ref{sec:VA} covers
the variational methodology and our proposed solutions. Section
\ref{sec:simul} reports simulation results and an application to the
real MRFM data. Section \ref{sec:concl} discusses our findings and
concludes.

\section{Formulation}
\label{sec:formulation}

\subsection{Image Model}

As in \citep{Park2012,Dobigeon2009a}, the image model is defined as:
    \begin{equation}
    \label{eq:model_nD}
    \Vobs = \MATtrans\Vima + \Vnoise = \ftrans{\psf}{\Vima} + \Vnoise,
    \end{equation}
where $\Vobs$ is a $P \times 1$ vectorized measurement, $\Vima =
[x_1,\ldots, x_N]^T \succeq 0$ is an $N \times 1$ vectorized sparse
image to be recovered, $\ftrans{\psf}{\cdot}$ is a convolution
operator with the PSF $\psf$, $\MATtrans = [\bfh_1 ,\ldots, \bfh_N]$
is an equivalent system matrix, and $\Vnoise$ is the measurement
noise vector. In this work, the noise vector $\Vnoise$  is assumed
to be Gaussian\footnote{$\calN(\boldsymbol{\mu},\boldsymbol{\Sigma})$
denotes a Gaussian random variable with mean $\boldsymbol{\mu}$ and covariance matrix
$\boldsymbol{\Sigma}$.}, $\Vnoise \sim
\calN\left(\Vzero,\noisevar\Id{\dimn}\right)$. The PSF $\psf$ is assumed to be unknown but a
nominal PSF estimate $\psf_0$ is available. The semi-blind
deconvolution problem addressed in this paper consists of the joint
estimation of $\Vima$ and $\psf$ from the noisy measurements
$\Vobs$ and nominal PSF $\psf_0$.

\subsection{PSF Basis Expansion}
\label{ssec:PSFbases}

The nominal PSF $\psf_0$ is assumed to be generated with known parameters (gathered in the vector $\boldsymbol{\zeta_0}$) tuned during imaging experiments. However, due
to model mismatch and experimental errors, the true PSF $\psf$ may deviate from the nominal PSF $\psf_0$.
If the generation model for PSFs is complex, direct estimation of a parameter deviation, $\Delta\boldsymbol{\zeta} = \boldsymbol{\zeta_{true}} - \boldsymbol{\zeta_0}$, is difficult.

We model the PSF $\psf$ (resp. $\{\bfH\}$) as a perturbation about a nominal PSF $\psf_0$ (resp. $\{\bfH^0\}$) with $K$ basis vectors $\psf_k$, $k=1,\ldots,K$, that span a subspace
representing possible perturbations $\Delta\psf$. We empirically determined this basis using the following PSF
variational eigendecomposition approach. A number of PSFs $\tilde\psf$ are
generated following the PSF generation model with parameters $\boldsymbol{\zeta}$ randomly
drawn according to Gaussian distribution\footnote{ The variances of the Gaussian distributions are carefully tuned so that their standard deviations
produce a minimal volume ellipsoid that contains the set of valid PSFs. }
centered at the nominal values $\boldsymbol{\zeta_0}$. Then a standard principal component analysis (PCA) of the
residuals $\left\{\tilde\psf_j-\psf_0\right\}_{j=1,\ldots}$
is used to identify $K$ principal axes that are associated with the basis vectors
$\psf_k$. The necessary number of basis vectors, $K$, is determined empirically by detecting a knee at the
scree plot. The first few eigenfunctions, corresponding to the first few largest
eigenvalues, explain major portion of the observed perturbations.
\red{If there is no PSF generation model, then we 
can decompose the support region of the true (suspected) PSF to produce an orthonormal basis.
The necessary number of the bases is again chosen to 
 explain most support areas that have major portion/energy of the desired PSF.
 This approach is presented in our experiment with Gaussian PSFs.}

We use a basis expansion to present $\psf(\bfc)$ as the following linear approximation to $\psf$,
\begin{equation}
 \psf(\bfc) = \psf_0 + \sum_{i=1}^K c_i \psf_i ,
\end{equation}
where the $\{ c_i\}$ determine the PSF relative to this bases.
With this parameterization, the objective of semi-blind
deconvolution is to estimate the unknown image, $\bfx$, and the
linear expansion coefficients $\bfc = [c_1 ,\ldots, c_K]^T$.

\subsection{Determination of Priors}
\label{ssec:priors}
The priors on the PSF, the image, and the noise are constructed as
latent variables in a hierarchical
Bayesian model.

\subsubsection{Likelihood function}
Under the hypothesis that the noise in \eqref{eq:model_nD} is white
Gaussian, the likelihood function takes the form
\begin{align}
  \label{eq:likelihood}
  p\left(\Vobs | \Vima,\bfc,\noisevar \right) = &
  \left(\frac{1}{2\pi\sigma^2}\right)^{\frac{\dimn}{2}} \times \notag \\
&   \exp\left(-\frac{\norm{\Vobs-\ftrans{\psf\left(\bfc\right)}{\Vima}}^2}{2\noisevar}\right),
\end{align}
where $\norm{\cdot}$ denotes the $\ell_2$ norm
$\norm{\bfx}^2=\bfx\transp\bfx$.

\subsubsection{Image and label priors}
\label{subsubsec:prior_ima}
To induce sparsity and positivity of the image, we use an image prior consisting of  ``a mixture
of a point mass at zero and a single-sided exponential distribution'' \citep{Ting2009,Dobigeon2009a,Park2012}. This prior
is a convex combination of an atom at zero
and an exponential distribution:
\begin{equation}
    \label{eq:LAZE1}
  p(x_i|a,w) = (1-w)\delta(x_i) + w g(x_i|a) .
\end{equation}

In \eqref{eq:LAZE1}, $\delta(\cdot)$ is the Dirac delta function, $w =
\mathrm{P}\left(x_i\neq 0\right)$ is the prior probability of a
non-zero pixel and
$g(x_i|a)=\frac{1}{a}\exp\left(-\frac{x_i}{a}\right)
\boldsymbol{1}_{\mathbb{R}^*_+}(x_i)$ is a single-sided exponential
distribution where $\mathbb{R}^*_+$ is a set of positive real numbers and $\boldsymbol{1}_{\mathbb{E}}(\cdot)$ denotes the indicator
function on the set $\mathbb{E}$:
\begin{equation}
  \boldsymbol{1}_{\mathbb{E}}(x) = \left\{
                                     \begin{array}{ll}
                                       1, & \hbox{if $x\in\mathbb{E}$;} \\
                                       0, & \hbox{otherwise.}
                                     \end{array}
                                   \right.
\end{equation}

A distinctive property of the image prior \eqref{eq:LAZE1} is
that it can be expressed as a latent variable model
\begin{equation} \label{eq:p_x_bar_az}
p(x_i|a,z_i) = (1-z_i)\delta(x_i) + z_i g(x_i|a) ,
\end{equation}
where the binary variables $\{ z_i \}_1^N$ are independent and identically distributed and indicate
if the pixel $x_i$ is active
\begin{equation}
  z_i = \left\{ \begin{array}{ll}
                                       1, & \hbox{if $x_i\neq 0$;} \\
                                       0, & \hbox{otherwise.}
                                     \end{array}
                                   \right.
\end{equation}
and have the Bernoulli probabilities: $z_i \sim Ber(w)$.

The prior
distribution of pixel value $x_i$ in \eqref{eq:LAZE1} can be
rewritten conditionally upon latent variable $z_i$
\begin{align*}
  p\left(x_i|z_i=0\right)  &= \delta\left(x_i\right),\\
  p\left(x_i|a,z_i=1\right)       &= g\left(x_i|a\right),
\end{align*}
which can be summarized in the following factorized form
\begin{equation}
  p(x_i|a,z_i)=\delta(x_i)^{1-z_i}g(x_i|a)^{z_i}.
\end{equation}
By assuming each component $x_i$ to be
conditionally independent given $z_i$ and $a$, the following
conditional prior distribution is obtained for $\mathbf{x}$:
\begin{equation}
  p(\mathbf{x}|a,\mathbf{z}) = \prod_{i=1}^{N} \left[\delta(x_i)^{1-z_i}g(x_i|a)^{z_i}\right]
\end{equation}
where $\mathbf{z}=\left[z_1,\ldots,z_N\right]$.

This factorized form will turn out to be crucial for simplifying
the variational Bayes reconstruction algorithm in Section \ref{sec:VA}.

\subsubsection{PSF parameter prior}
We assume that the PSF parameters $c_{1},\ldots,c_{K}$ are independent
and $c_k$ is uniformly distributed over intervals
\begin{equation}
  \calS_k
=\left[-\Delta c_{k},\Delta c_{k}\right] .
\end{equation}
These intervals are specified a priori and are
associated with error tolerances of the imaging instrument. The joint
prior distribution of $\bfc=\left[c_{1},\ldots,c_{K}\right]\transp$ is therefore:
\begin{equation}
  p\left(\bfc\right) = \prod_{k=1}^{K}\frac{1}{2\Delta c_{k}}
\Indicfun{\calS_k}{c_{k}}.
\end{equation}

\subsubsection{Noise variance prior} A conjugate
inverse-Gamma distribution with parameters $\varsigma_0$ and
$\varsigma_1$ is assumed as the prior distribution for the noise
variance (See \ref{appen:IG} for the details of this distribution):
\begin{equation}
    \label{eq:prior_noisevar}
  \noisevar | \varsigma_0, \varsigma_1 \sim
\calI\calG\left(\varsigma_0,\varsigma_1\right).
\end{equation}
The parameters $\varsigma_0$ and $\varsigma_1$ will be fixed to a
number small enough to obtain a vague hyperprior, unless we have good prior knowledge.

\subsection{Hyperparameter Priors}
As reported in \citep{Ting2009,Dobigeon2009a}, the values of
the hyperparameters $\left\{a,w\right\}$ greatly impact the quality
of the deconvolution. Following the approach in
 \citep{Park2012}, we propose to include them within the
Bayesian model, leading to a second level of hierarchy in the
Bayesian paradigm. This hierarchical Bayesian model requires the
definition of prior distributions for these hyperparameters, also
referred to as \emph{hyperpriors} which are defined below.

\subsubsection{Hyperparameter $a$} A conjugate inverse-Gamma
distribution is assumed for the Laplacian scale parameter $a$:
\begin{equation}
  a | \Valpha \sim \calI\calG\left(\alpha_0,\alpha_1\right),
\end{equation}
with $\Valpha=\left[\alpha_0,\alpha_1\right]\transp$.
The parameters $\alpha_0$ and $\alpha_1$ will be fixed to a
number small enough to obtain a vague hyperprior, unless we have good prior knowledge.

\subsubsection{Hyperparameter $w$}\label{sssec:w}
We assume a Beta random variable with parameters $(\beta_0, \beta_1)$, which are iteratively updated in accordance with data fidelity.
The parameter values will reflect the degree of prior knowledge and we set $\beta_0 = \beta_1 =1$ to obtain a non-informative prior. (See \ref{appen:Beta} for the details of this distribution)
\begin{equation}
  w   \sim  \calB(\beta_0,\beta_1).
\end{equation}

\subsection{Posterior Distribution}
\begin{figure}
\begin{center}
\includegraphics[width=0.3\textwidth]{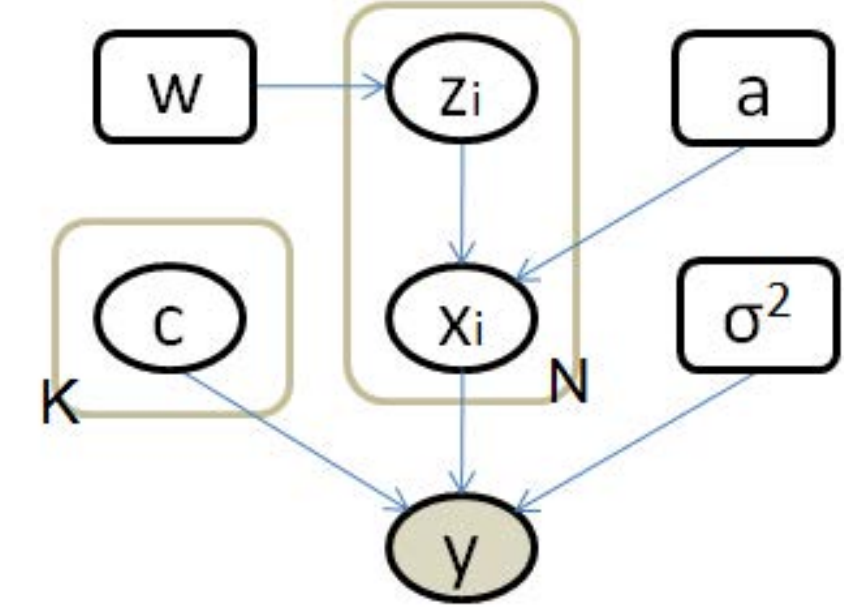}
\end{center}
\caption{Conditional relationships between variables. A node at an
arrow tail conditions the node at the arrow head.} \label{fig:GM}
\end{figure}

The conditional relationships between variables is illustrated
in Fig.~\ref{fig:GM}. The resulting posterior of hidden variables
given the observation is
\begin{multline}
  p( \bfx, a, \bfz, w, \bfc, \sigma^2 | \bfy ) \propto
p(\bfy|\bfx,\bfc,\sigma^2) \\ \times p(\bfx|a,\bfz) p(\bfz|w) p(w)
p(a) p(\bfc) p(\sigma^2) .
\end{multline}
Since it is too complex to derive exact Bayesian estimators from
this posterior, a variational approximation of this distribution is
proposed in the next section.

\section{Variational Approximation}
\label{sec:VA}

\subsection{Basics of Variational Inference}
In this section, we show how to approximate the posterior densities
within a variational Bayes framework.
Denote by $\bfU$ the set of all hidden parameter variables including the image variable $\bfx$ in the model, denoted by $\calM$.
The hierarchical model implies the Markov representation $p(\bfy,\bfU|\calM) = p(\bfy|\bfU,\calM) p(\bfU|\calM) $. Our objective is to compute the posterior $p(\bfx|\bfy, \calM) = \int p(\bfy|\bfU,\calM)  p(\bfU|\calM) d \bfU_{\backslash \bfx} / p(\bfy|\calM)$, where $\bfU_{\backslash \bfx}$ is a set of variables in $\bfU$ except $\bfx$.
Let $q$ be any arbitrary distribution of $\bfU$. Then
\begin{equation}
\ln p(\bfy|\calM) = \calL(q) + \mathrm{KL}( q \| p)
\end{equation}
with
\begin{align}
  \calL(q) &= \int q(\bfU|\calM) \ln\left(\frac{p(\bfy, \bfU|\calM)}{q(\bfU|\calM)}\right)  d\bfU\\
    \mathrm{KL}( q \| p) &= - \int q(\bfU|\calM) \ln\left(\frac{p(\bfU|\bfy,
    \calM)}{q(\bfU|\calM)}\right) d\bfU .
\end{align}

We observe that maximizing the lower bound
$\calL(q)$ is equivalent to
minimizing the Kullback-Leibler (KL) divergence
$\mathrm{KL}( q \| p) $.
Consequently, instead of directly evaluating $p(\bfy|\calM)$ given $\calM$, we will specify a
distribution $q(\bfU|\calM)$ that approximates the posterior $p(\bfU | \bfy,\calM)$.
The best approximation maximizes $\calL(q)$.
We present Algorithm~\ref{algo:VAiter} that iteratively increases the value of $\calL(q)$ by updating posterior surrogate densities.
To obtain a tractable
approximating distribution $q$, we will assume a
factorized form as $q(\bfU) = \prod_j q(\bfU_j)$ where $\bfU$ has been
partitioned into disjoint groups $\bfU_j$.
Subject to this factorization constraint, the optimal distribution
$q^*\left(\bfU\right)= \prod_j q^*(\bfU_j)$ is given by
\begin{equation}
\label{eq:KL} \ln q^*_j(\bfU_j) = \EE_{\backslash \bfU_j} \left[\ln p(\bfU, \bfy) \right]
+ (\mathrm{const}), \quad \forall j
\end{equation}
where $\EE_{\backslash \bfU_j}$ denotes the expectation\footnote{In the
sequel, we use both $\EE\left[ \cdot \right] $ and
$\langle \cdot \rangle$ to denote the expectation. To make our expressions more compact, we use subscripts to denote expectation with respect to the random variables
in the subscripts.
These notations with the subscripts of `$\backslash \bfv$' denote expectation
with respect to all random variables except for the variable $\bfv$. e.g. $\EE_{\backslash \bfU_j}$
}
with respect to all factors
$\bfU_i$ except $i = j$.
We will call $q^*(\bfU)$ the posterior surrogate for $p$.

\subsection{Suggested Factorization}
\label{ssec:suggestedFactorization}

Based on our assumptions on the image and hidden parameters, the
random vector is $\bfU \triangleq
\left\{\boldsymbol{\theta},\boldsymbol{\phi}\right\} = \{\bfx, a, \bfz,
 w,\bfc, \sigma^2\}$ with $\boldsymbol{\theta}=\left\{\bfx, \bfz,
\bfc\right\}$ and $\boldsymbol{\phi}=\left\{a,w,  \sigma^2\right\}$.
We propose the following factorized approximating distribution
\begin{equation}
\label{eq:q_U}
q(\bfU) = q(\bfx, a, \bfz, w,\bfc, \sigma^2) = q(\bfx,\bfz, \bfc) q(a,w,  \sigma^2).
\end{equation}
Ignoring constants\footnote{In the sequel, constant terms with respect to the variables of interest can be omitted in equations.}, \eqref{eq:KL} leads to
\begin{align}
   \ln q( a ,w,\sigma^2) &= \underbrace{ \EE_{\backslash a} \ln p(\bfx|a,\bfz)p(a)}_{\ln q(a)}  +  \notag\\
                         &  \underbrace{ \EE_{\backslash w} \ln p(\bfz|w)p(w) }_{\ln q(w)} + \underbrace{ \EE_{\backslash \sigma^2} \ln p(\bfy|\bfx,\sigma^2) p(\sigma^2)}_{\ln q(\sigma^2)}
\end{align}
which induces the factorization
\begin{equation}
  q(\boldsymbol{\phi}) = q(a)q(w)q(\sigma^2).
\end{equation}
Similarly, the factorized distribution for $\bfx$, $\bfz$ and $\bfc$
is
\begin{equation}
  q\left(\boldsymbol{\theta}\right) = \left[\prod_i q(x_i|z_i)\right] q(\bfz) q(\bfc) 
\end{equation}
leading to the fully factorized distribution
\begin{equation}
\label{eq:factorization}
  q\left(\boldsymbol{\theta},\boldsymbol{\phi}\right) =\left[\prod_{i} q(x_i|z_i)\right]  q(a) q(\bfz) q(w) q(\bfc) q(\sigma^2)
\end{equation}

\subsection{Approximating Distribution $q$}
\label{ssec:solutions}
In this section, we specify the marginal distributions in the
approximated posterior distribution required in
\eqref{eq:factorization}. More details are
described in \ref{appen:deriv_q}. The parameters for the
posterior distributions are evaluated iteratively due to the mutual
dependence of the parameters in the distributions for the hidden
variables, as illustrated in Algorithm~\ref{algo:VAiter}.

\subsubsection{Posterior surrogate for $a$}
\begin{equation}\label{q_a}
q(a) = \calI\calG(\tilde\alpha_0, \tilde\alpha_1 ),
\end{equation}
with $\tilde\alpha_0 = \alpha_0+ \sum \langle z_i \rangle, \tilde\alpha_1 = \alpha_1 + \sum \langle z_i x_i \rangle $.

\subsubsection{Posterior surrogate for $w$}
\begin{equation}\label{q_w}
q(w) = \calB(\tilde\beta_0,\tilde\beta_1),
\end{equation}
with $\tilde\beta_0 =  \beta_0+N-\sum \langle z_i \rangle, \tilde\beta_1 = \beta_1 + \sum \langle z_i \rangle$.

\subsubsection{Posterior surrogate for $\sigma^2$}
\begin{equation}\label{q_sigma2}
q(\sigma^2) = \calI\calG(\tilde\varsigma_0, \tilde\varsigma_1 ),
\end{equation}
with $ \tilde\varsigma_0 = P/2 + \varsigma_0$, $\tilde\varsigma_1 =
\langle \|\bfy-\bfH\bfx\|^2 \rangle /2 + \varsigma_1$,  and $\langle
\|\bfy-\bfH\bfx\|^2 \rangle = \|\bfy - \langle \bfH \rangle \langle
\bfx \rangle \|^2 + \sum \mathrm{var}[x_i] \left[ \| \langle \psf
\rangle \|^2 + \sum_l \sigma_{c_l} \|\psf_l\|^2\right] + \sum_l
\sigma_{c_l} \|{\bfH^l\langle\bfx\rangle}\|^2$,
 where $\sigma_{c_l}$ is the variance of the Gaussian distribution $q(c_l)$
 given in \eqref{q_c} and $\mathrm{var}[x_i]$ is computed
under the distribution $q(x_i)$ defined in the next section and
described in  \ref{apssec:q_x}.

\subsubsection{Posterior surrogate for $\mathbf{x}$}
\label{sssec:q_x}
We first note that
\begin{equation}
   \ln q(\bfx,\bfz) = \ln q(\bfx|\bfz) q(\bfz) = \EE \left[ \ln
p(\bfy|\bfx,\sigma^2)p(\bfx|a,\bfz)p(\bfz|w) \right] .
\end{equation}
The conditional density of $\bfx$ given $\bfz$ is $p(\bfx|a,\bfz) =
\prod_i^N g_{z_i} (x_i)$, where $g_0(x_i) \triangleq \delta(x_i),
g_1(x_i) \triangleq g(x_i|a)$.  Therefore, the conditional posterior surrogate for $x_i$ is
\begin{align}\label{q_x_z}
q(x_i|z_i = 0) &= \delta(x_i),  \\
q(x_i|z_i = 1) &= \phi_+(\mu_i,\eta_i),
\end{align}
where $\phi_+(\mu,\sigma^2)$ is a positively truncated-Gaussian
density function with the hidden mean $\mu$ and variance $\sigma^2$,
$\eta_i = 1/[\langle\|\bfh_i\|^2 \rangle \langle 1/\sigma^2
\rangle]$, $\mu_i = \eta_i[\langle \bfh_i^T \bfe_i \rangle
\langle 1/\sigma^2 \rangle -  \langle 1/a \rangle]$,
$\bfe_i = \bfy - \bfH \bfx_{-i}$,  $\bfx_{-i}$ is $\bfx$ except for the $i$th entry replaced with 0, and $\bfh_i$ is the $i$th column of $\bfH$. Therefore,
\begin{equation}\label{q_x}
q(x_i) = q(z_i =0) \delta(x_i) + q(z_i=1) \phi_+(\mu_i,\eta_i),
\end{equation}
which is a Bernoulli truncated-Gaussian density.

\subsubsection{Posterior surrogate for $\mathbf{z}$}
For $i=1,\ldots,N$,
\begin{align}\label{q_z}
q(z_i=1) = {1}/[{1+C^\prime_i}]  \text{ and }
q(z_i=0) = 1-q(z_i=1),
\end{align}
with $C^\prime_i = \exp(C_i/2 \times
\tilde\varsigma_0/\tilde\varsigma_1 + \mu_i
\tilde\alpha_0/\tilde\alpha_1 + \ln \tilde\alpha_1 -
\psi(\tilde\alpha_0) + \psi(\tilde\beta_0) - \psi(\tilde\beta_1) )$.
$\psi$ is the digamma function and $C_i = \langle \|\bfh_i\|^2
\rangle (\mu_i^2 + \eta_i) -2 \langle \bfe_i^T\bfh_i \rangle \mu_i$.

\subsubsection{Posterior surrogate for $\mathbf{c}$}\label{sssec:q_c}
For $j = 1, \ldots, K$,
\begin{equation}\label{q_c}
q(c_j) = \phi(\mu_{c_j}, \sigma_{c_j}),
\end{equation}
where $\phi(\mu,\sigma)$ is the probability density function for the
normal distribution with the mean $\mu$ and variance $\sigma$,
$\mu_{c_j} = \dfrac{ \langle \bfx^T {\bfH^j}^T \bfy - \bfx {\bfH^j}^T
\bfH^0 \bfx - \sum_{l \neq j} \bfx^T {\bfH^j}^T \bfH^l c_l \bfx
\rangle }{ \langle \bfx^T {\bfH^j}^T \bfH^j \bfx \rangle }$, and
$1/\sigma_{c_j} = \langle 1/\sigma^2 \rangle \langle \bfx^T
{\bfH^j}^T \bfH^j \bfx \rangle $.

\begin{algorithm}[h]
\caption{VB semi-blind image reconstruction algorithm}\label{algo:VAiter}
\begin{algorithmic}[1]
    \STATE \emph{\scriptsize{\% Initialization:}}
    \STATE Initialize estimates $\langle\sample{\Vima}{0}\rangle$, $\langle \bfz^{(0)}\rangle$, and $w^{(0)}$, and set $\bfc = \bf 0$ to have $\hat\psf^{(0)} = \psf_0$,
    \STATE \emph{\scriptsize{\% Iterations:}}
    \FOR{$t=1,2, \ldots, $}
                \STATE Evaluate $\tilde\alpha_0^{(t)}, \tilde\alpha_1^{(t)}$ in \eqref{q_a} by using $\langle \Vima^{(t-1)} \rangle, \langle \bfz^{(t-1)} \rangle$,
        \STATE Evaluate $\tilde\beta_0^{(t)}, \tilde\beta_1^{(t)}$ in \eqref{q_w} by using $\langle \bfz^{(t-1)} \rangle$,
        \STATE Evaluate $\tilde\varsigma_0^{(t)}, \tilde\varsigma_1^{(t)}$ in \eqref{q_sigma2}  from $\langle\|\bfy-\bfH \bfx\|^2\rangle$,
        \FOR{$i=1,2, \ldots,N $}
        \STATE Evaluate necessary statistics ($\mu_i, \eta_i$) for $q(x_i| z_i =1)$ in \eqref{q_x_z},
        \STATE Evaluate $q(z_i =1)$ in \eqref{q_z},
        \STATE Evaluate $\langle x_i \rangle, \mathrm{var}[x_i]$,
        \STATE For $l=1,\ldots,K$, evaluate $\mu_{c_l}, 1/\sigma_{c_l}$ for $q(c_l)$ in \eqref{q_c},
                \ENDFOR
    \ENDFOR
\end{algorithmic}
\end{algorithm}

The final iterative algorithm is presented in
Algorithm~\ref{algo:VAiter}, where required \emph{shaping}
parameters under distributional assumptions and related statistics
are iteratively updated.

\section{Simulation Results}
\label{sec:simul} We first present numerical results obtained for
Gaussian and typical MRFM PSFs, shown in
Fig.~\ref{fig:myopic_x11_Gaussian_figTrueObsPSFs} and
Fig.~\ref{fig:myopic_x11_MRFM_figTrueObsPSFs}, respectively. Then
the proposed variational algorithm is applied to a tobacco virus MRFM
data set.
There are many possible approaches to selecting hyperparameters, including the non-informative approach of \citep{Park2012} and the expectation-maximization approach of \citep{Nasios2006}.
In our experiments, hyper-parameters
$\varsigma_0$, $\varsigma_1$, $\alpha_0$, and $\alpha_1$
for the densities
are chosen based on the framework advocated in \citep{Park2012}. This leads to the vague priors corresponding to selecting small values $\varsigma_0 = \varsigma_1 = \alpha_0=\alpha_1=1$.
For $w$, the noninformative  initialization is made by setting $\beta_0 = \beta_1 = 1$, which gives flexibility to the surrogate posterior density for $w$.
The resulting prior Beta distribution for $w$ is a uniform distribution on $[0,1]$ for the mean proportion of non-zero pixels.
\begin{equation}
  w   \sim  \calB(\beta_0,\beta_1) \sim \calU\left([0,1]\right).
\end{equation}

The initial image used to initialize the algorithm is
obtained from one Landweber iteration \citep{Landweber1951}.

\subsection{Simulation with Gaussian PSF}
\label{ssec:sim_Gaussian}

\begin{figure}[h]
  \centering
 \subfigure[True image $\bfx$]{\label{fig:2a}
  \includegraphics[width=\figwidthsmall]{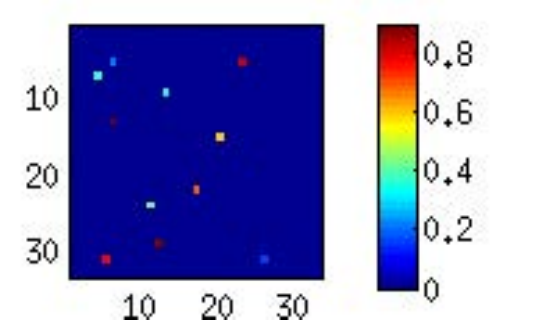}}
 \subfigure[Obsevation]{\label{fig:2b}
  \includegraphics[width=\figwidthsmall]{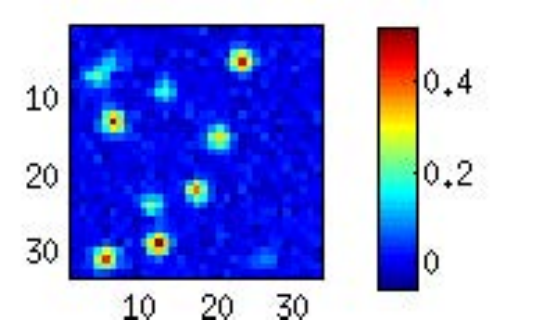}}
 \subfigure[True PSF ]{\label{fig:2c}
  \includegraphics[width=\figwidthsmall]{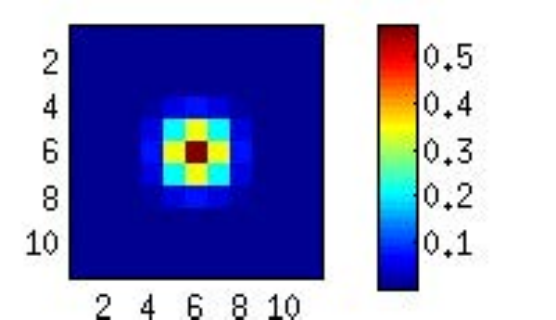}}
 \subfigure[Mismatched PSF ]{\label{fig:2d}
  \includegraphics[width=\figwidthsmall]{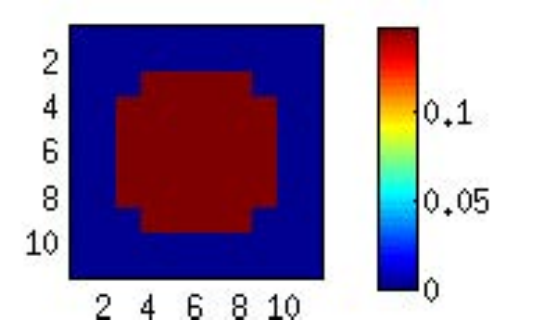}}
  \caption{Experiment with Gaussian PSF: true image, observation, true PSF, and mismatched PSF ($\psf_0$).}
  \label{fig:myopic_x11_Gaussian_figTrueObsPSFs}
\end{figure}

\begin{figure}[h!]
 \centering
 \subfigure[ $\log\|\bfy - \EE\bfH\EE\bfx\|^2$ (solid line) and noise level (dashed line)]{\label{fig:3a}
  \includegraphics[width=\figwidthFigThree]{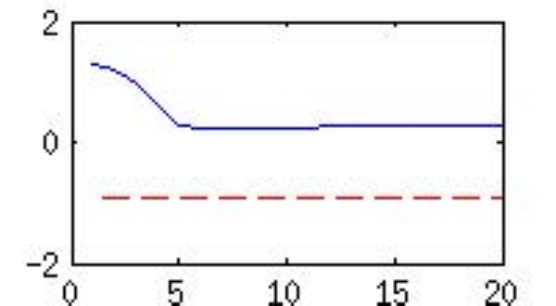}}
 \subfigure[ $\log\| \bfx_{true} -\EE\bfx \|^2$]{\label{fig:3b}
  \includegraphics[width=\figwidthFigThree]{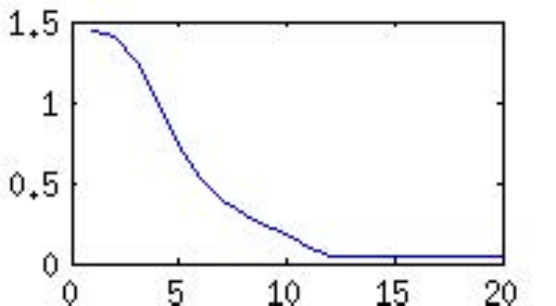}}
 \subfigure[$\EE {\left[ 1/a \right]}$  (solid line) and true value (dashed line) ]{\label{fig:3c}
  \includegraphics[width=\figwidthFigThree]{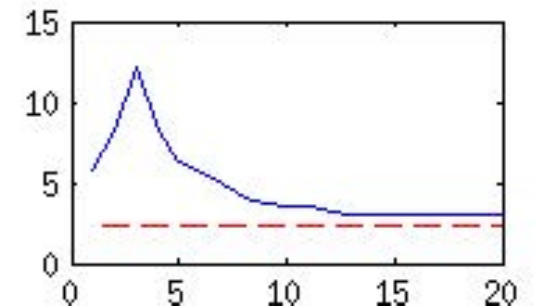}}
 \subfigure[$\EE {\left[ 1/\sigma^2\right]}$  (solid line) and true value (dashed line)  ]{\label{fig:3d}
  \includegraphics[width=\figwidthFigThree]{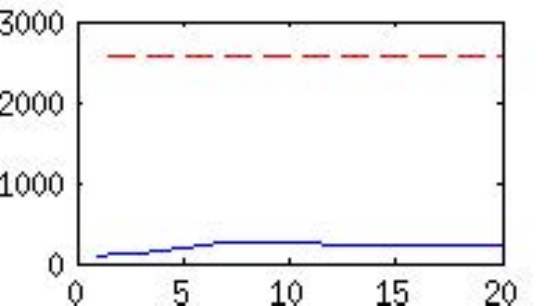}}
 \subfigure[$\EE {\left[ w \right]}$ (solid line) and true value (dashed line) ]{\label{fig:3e}
  \includegraphics[width=\figwidthFigThree]{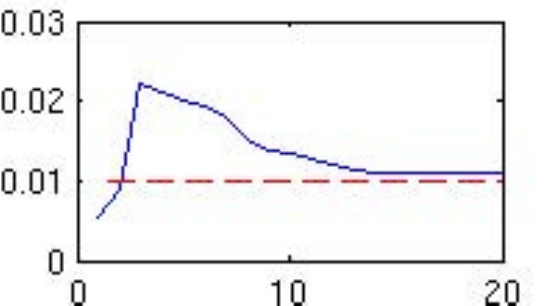}}
 \subfigure[$\EE {\left[ \bfc\right]}$. Four PSF coefficients. ]{\label{fig:3f}
  \includegraphics[width=\figwidthFigThree]{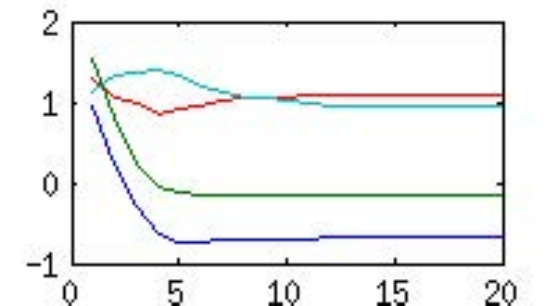}}
 \caption{Result of Algorithm~\ref{algo:VAiter}: curves of residual, error, $\EE\left[1/a\right], \EE\left[1/\sigma^2\right], \EE\left[w\right], \EE\left[\bfc\right]$, as functions of number of iterations. These curves show
    how fast the convergence is achieved. }
 \label{fig:myopic_x11_Gaussian_figCurves}
\end{figure}

\begin{figure}[h]
  \centering
 \subfigure[Estimated PSF]{\label{fig:4a}
  \includegraphics[width=\figwidthsmall]{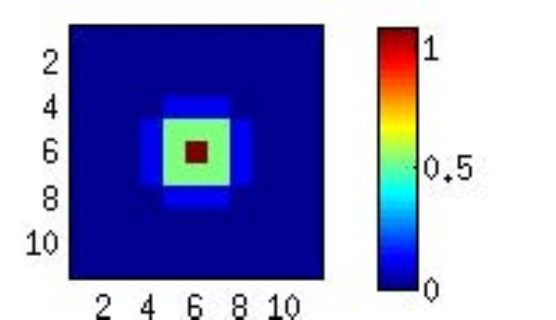}}
 \subfigure[Estimated image]{\label{fig:4b}
  \includegraphics[width=\figwidthsmall]{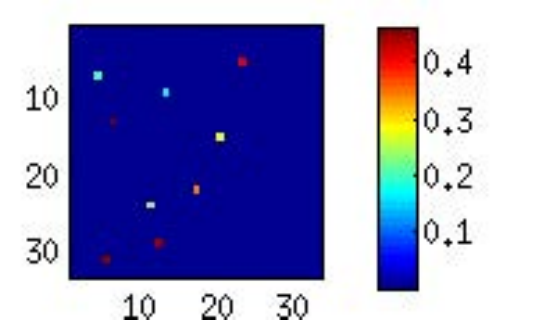}}
 \subfigure[Variance map]{\label{fig:4c}
  \includegraphics[width=\figwidthsmall]{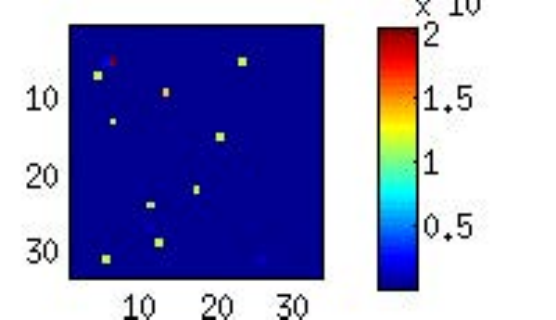}}
 \subfigure[Weight map]{\label{fig:4d}
  \includegraphics[width=\figwidthsmall]{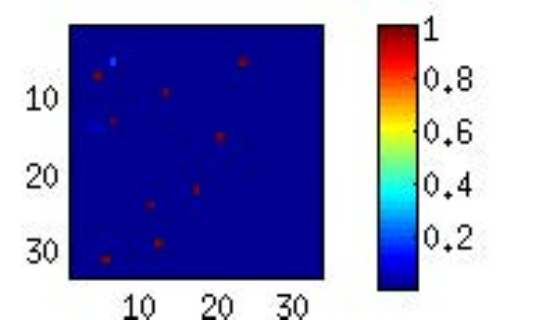}}
  \caption{ (a) Restored PSF, (b) image, (c) map of pixel-wise (posterior) variance, and (d) weight map. $\hat\psf=\EE\psf$ is close to the true one. A pixel-wise weight shown in (d) is the posterior probability of the pixel  being a nonzero signal.}
    \label{fig:myopic_x11_Gaussian_figImage}
\end{figure}

\begin{figure}[h]
  \centering
 \subfigure[The first basis $\psf_1$]{\label{fig:5a}
  \includegraphics[width=\figwidthsmall]{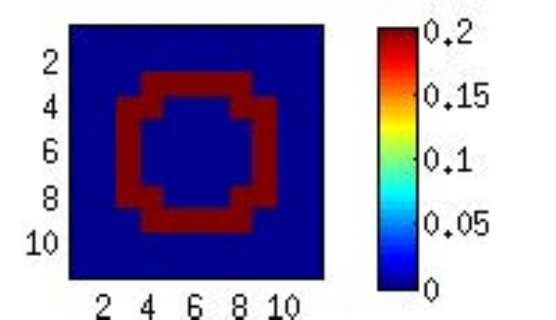}}
 \subfigure[The second basis $\psf_2$]{\label{fig:5b}
  \includegraphics[width=\figwidthsmall]{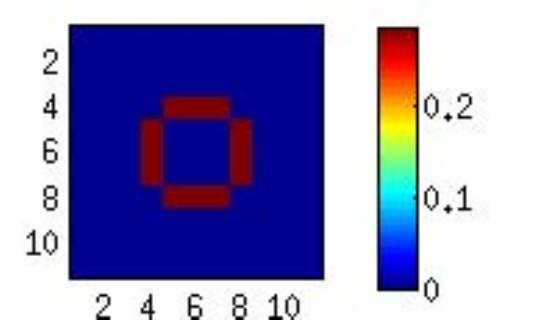}}
 \subfigure[The third basis $\psf_3$]{\label{fig:5c}
  \includegraphics[width=\figwidthsmall]{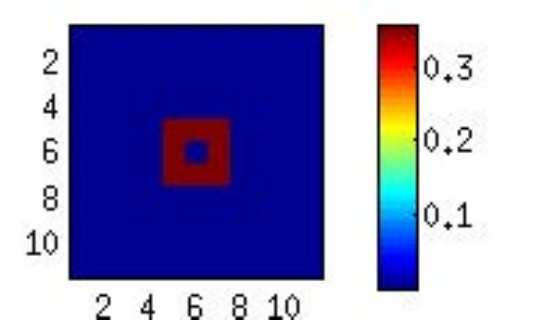}}
 \subfigure[The fourth basis $\psf_4$]{\label{fig:5d}
  \includegraphics[width=\figwidthsmall]{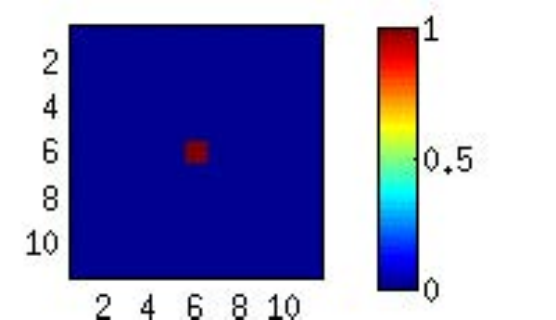}}
  \caption{PSF bases, $\psf_1,\ldots,\psf_4$, for Gaussian PSF.}
      \label{fig:myopic_x11_Gaussian_figKappas}
\end{figure}

The true image $\bfx$ used to generate the data, observation $\bfy$, the true PSF, and the initial, mismatched PSF are shown in
Fig.~\ref{fig:myopic_x11_Gaussian_figTrueObsPSFs}. Some quantities
of interest, computed from the outputs of the variational algorithm
are depicted as functions of the iteration number in
Fig.~\ref{fig:myopic_x11_Gaussian_figCurves}. These plots indicate
that convergence to the steady state is achieved after few iterations.
In Fig.~\ref{fig:myopic_x11_Gaussian_figCurves}, 
$\EE\left[w\right]$ and $\EE\left[1/a\right]$ get close to the true level but $\EE\left[1/\sigma^2\right]$ shows a deviation from the true values. \red{This large deviation implies that our estimation of noise level is conservative; the estimated noise level is larger  than the true level. This relates to the large deviation in projection error from noise level (Fig.~\ref{fig:3a}). } The drastic changes in the initial steps seen in the curves of $\EE\left[1/a\right], \EE\left[w\right]$ are due to the imperfect prior knowledge (initialization). 
 The final
estimated PSF and reconstructed image are depicted in
Fig.~\ref{fig:myopic_x11_Gaussian_figImage}, along with the reconstructed
variances and posterior probability of $z_i \neq 0$.
\red{
We decomposed the support region of the true PSF to produce orthonormal bases 
$\left\{\boldsymbol{\kappa}_i\right\}_i$ shown in Fig.~\ref{fig:myopic_x11_Gaussian_figKappas}. 
We extracted  4 bases because these four PSF bases clearly explain the significant part of the true Gaussian PSF. In other words, little energy resides outside of this basis set in PSF space.}

The reconstructed PSF clearly matches the true
one, as seen in Fig.~\ref{fig:myopic_x11_Gaussian_figTrueObsPSFs} and Fig.~\ref{fig:myopic_x11_Gaussian_figImage}. Note that
the restored image is slightly attenuated while the restored PSF is amplified
because of intrinsic scale ambiguity.

\subsection{Simulation with MRFM type PSFs}
\label{ssec:sim_MRFM}

\begin{figure}[h]
  \centering
 \subfigure[True image $\bfx$]{\label{fig:6a}
  \includegraphics[width=\figwidthsmall]{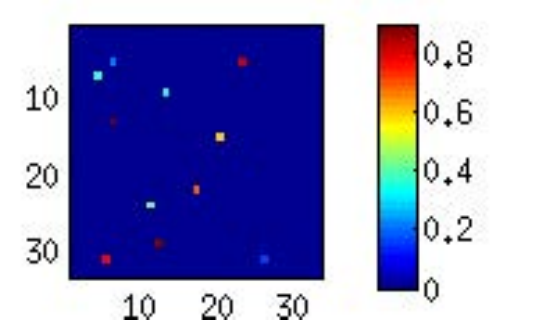}}
 \subfigure[Obsevation]{\label{fig:6b}
  \includegraphics[width=\figwidthsmall]{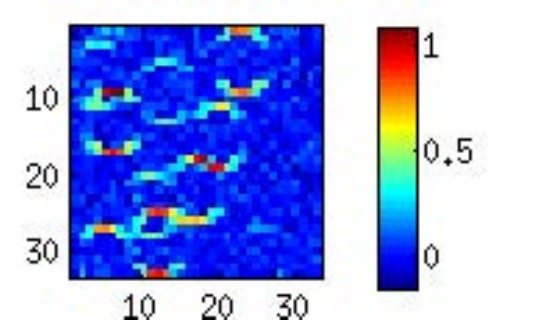}}
 \subfigure[True PSF ]{\label{fig:6c}
  \includegraphics[width=\figwidthsmall]{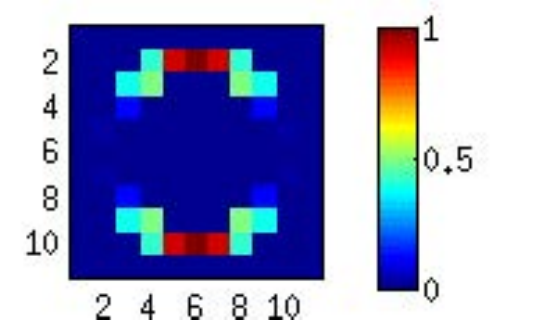}}
 \subfigure[Mismatched PSF ]{\label{fig:6d}
  \includegraphics[width=\figwidthsmall]{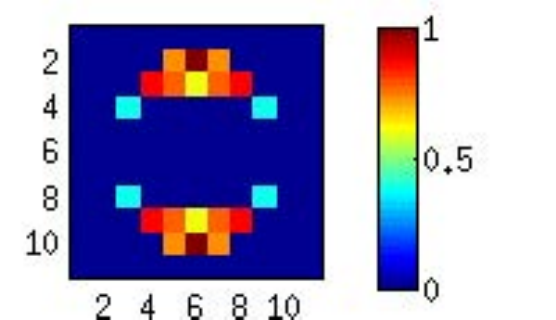}}
  \caption{Experiment with simplified MRFM PSF: true image, observation, true PSF, and mismatched PSF ($\psf_0$).}
    \label{fig:myopic_x11_MRFM_figTrueObsPSFs}
\end{figure}

\begin{figure}[h]
  \centering
 \subfigure[Estimated PSF]{\label{fig:7a}
  \includegraphics[width=\figwidthsmall]{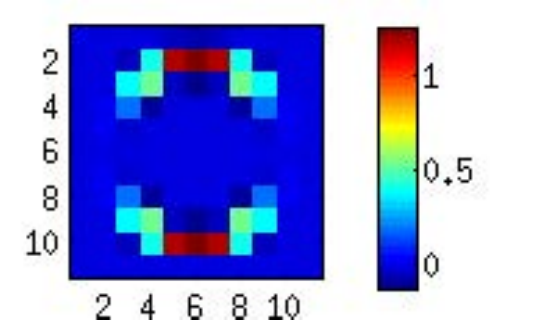}}
 \subfigure[Estimated image]{\label{fig:7b}
  \includegraphics[width=\figwidthsmall]{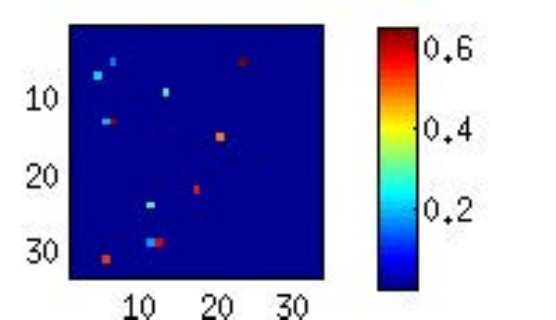}}
 \subfigure[Variance map]{\label{fig:7c}
  \includegraphics[width=\figwidthsmall]{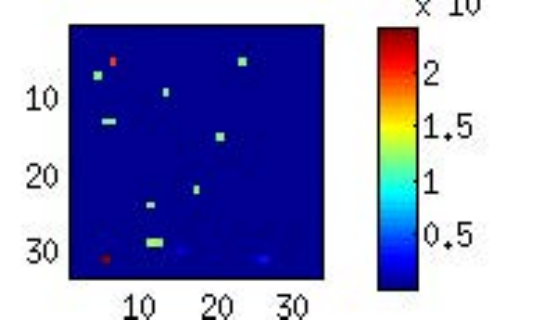}}
 \subfigure[Weight map]{\label{fig:7d}
  \includegraphics[width=\figwidthsmall]{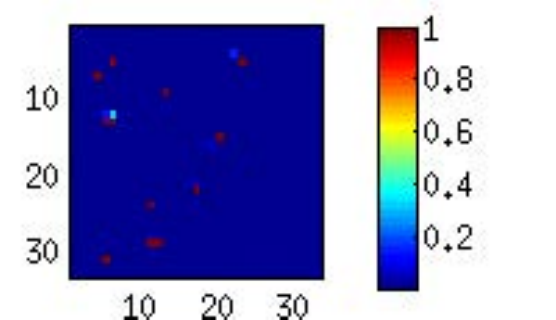}}
  \caption{Restored PSF and image with pixel-wise variance and weight map. $\hat\psf=\EE\psf$ is close to the true one.}
  \label{fig:myopic_x11_MRFM_figImage}
\end{figure}

\begin{figure}[h]
  \centering
 \subfigure[The first basis $\psf_1$]{\label{fig:8a}
  \includegraphics[width=\figwidthsmall]{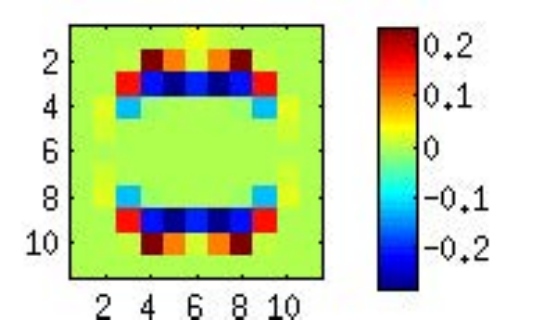}}
 \subfigure[The second basis $\psf_2$]{\label{fig:8b}
  \includegraphics[width=\figwidthsmall]{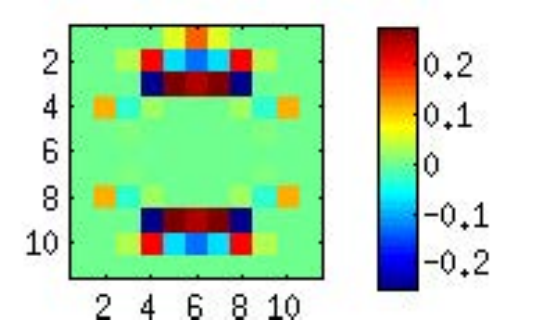}}
 \subfigure[The third basis $\psf_3$]{\label{fig:8c}
  \includegraphics[width=\figwidthsmall]{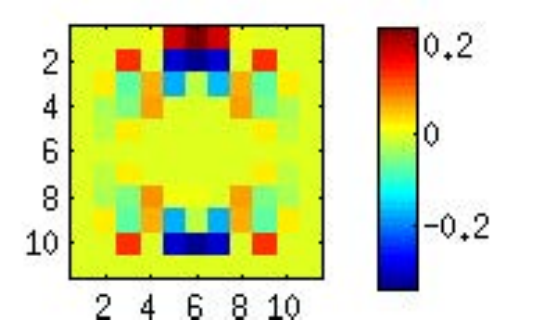}}
 \subfigure[The fourth basis $\psf_4$]{\label{fig:8d}
  \includegraphics[width=\figwidthsmall]{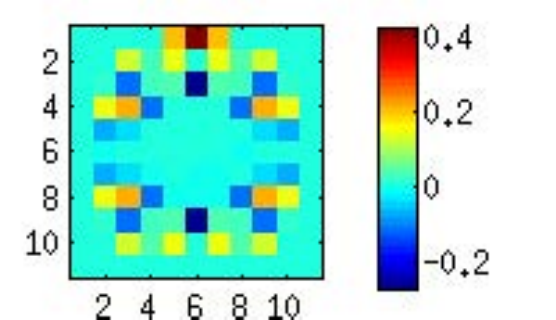}}
      \caption{PSF bases, $\psf_1,\ldots,\psf_4$, for MRFM PSF.}
    \label{fig:myopic_x11_MRFM_figKappas}
\end{figure}

The true image $\bfx$ used to generate the data, observation $\bfy$, the true PSF, and the initial, mismatched PSF are shown in
Fig.~\ref{fig:myopic_x11_MRFM_figTrueObsPSFs}.
The PSF models the PSF of the MRFM instrument, derived by
Mamin {\em et al.} \citep{Mamin2003}.
The convergence of the algorithm is achieved after the 10th iteration. The reconstructed image
can be compared to the true image in
Fig.~\ref{fig:myopic_x11_MRFM_figImage}, where the pixel-wise
variances and posterior probability of $z_i \neq 0$ are rendered.
The PSF bases are obtained by the procedure proposed in Section
\ref{ssec:PSFbases} with the simplified MRFM PSF model and the nominal parameter values \citep{Ting2009}.
\red{
Specifically, by detecting a knee $K = 4$ at the scree plot, explaining more than 98.69\% of the
observed perturbations (Fig.~3 in \citep{Park2012}), we use the first four eigenfunctions,
corresponding to the first four largest eigenvalues.
}
The resulting  $K = 4$ principal basis vectors are depicted in
Fig.~\ref{fig:myopic_x11_MRFM_figKappas}. The reconstructed PSF with
the bases clearly matches the true one, as seen in
Fig.~\ref{fig:myopic_x11_MRFM_figTrueObsPSFs} and Fig.~\ref{fig:myopic_x11_MRFM_figImage}.

\subsection{Comparison with PSF-mismatched reconstruction}

\begin{figure}[h]
  \centering
 \subfigure[True image]{\label{fig:9a}
  \includegraphics[width=\figwidthsmall]{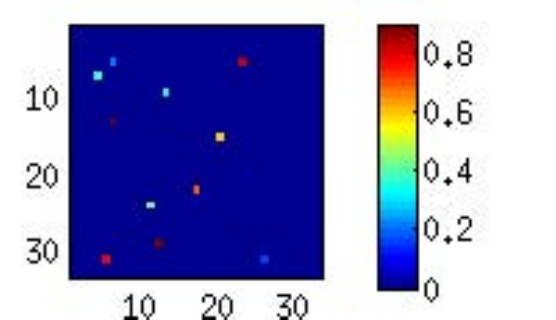}}
 \subfigure[Estimated image]{\label{fig:9b}
  \includegraphics[width=\figwidthsmall]{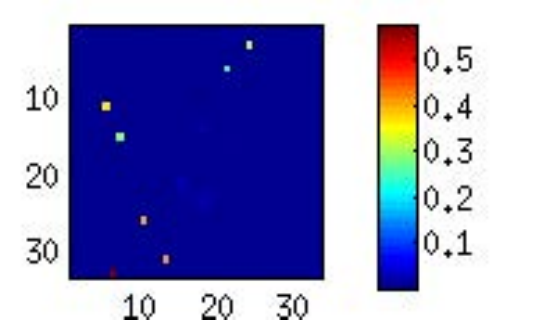}}
 \subfigure[Variance map]{\label{fig:9c}
  \includegraphics[width=\figwidthsmall]{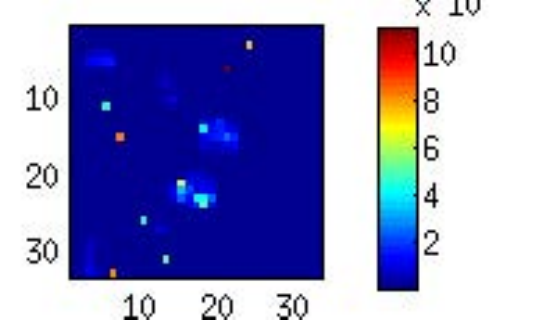}}
 \subfigure[Weight map]{\label{fig:9d}
  \includegraphics[width=\figwidthsmall]{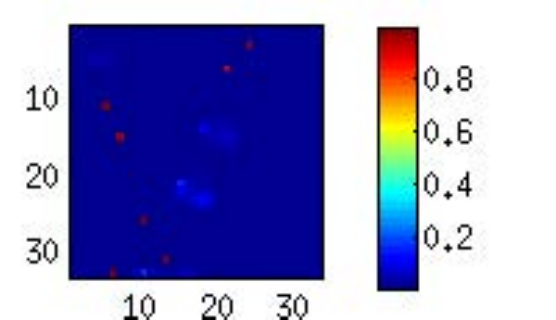}}
      \caption{(mismatched) Non-blind result with a mismatched Gaussian PSF.}
    \label{fig:NONmyopic_x11_Gaussian_figImage}
\end{figure}

\begin{figure}[h]
  \centering
 \subfigure[True image]{\label{fig:10a}
  \includegraphics[width=\figwidthsmall]{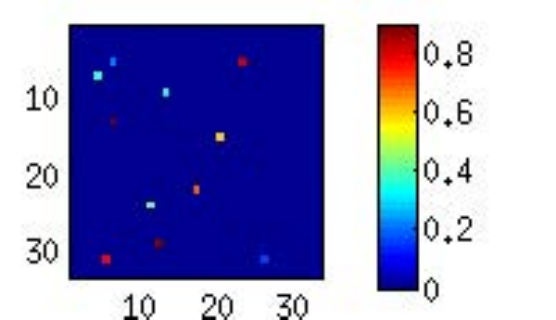}}
 \subfigure[Estimated image]{\label{fig:10b}
  \includegraphics[width=\figwidthsmall]{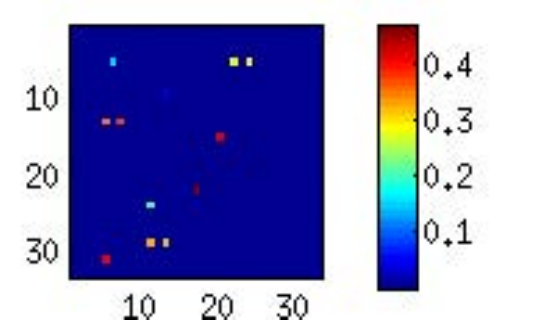}}
 \subfigure[Variance map]{\label{fig:10c}
  \includegraphics[width=\figwidthsmall]{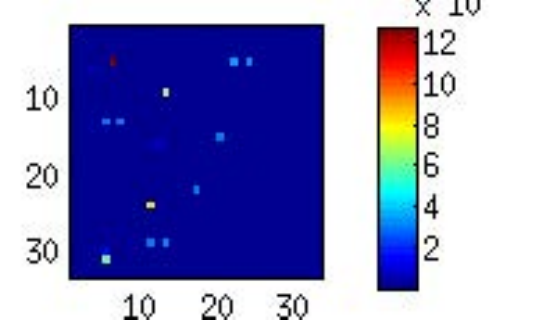}}
 \subfigure[Weight map]{\label{fig:10d}
  \includegraphics[width=\figwidthsmall]{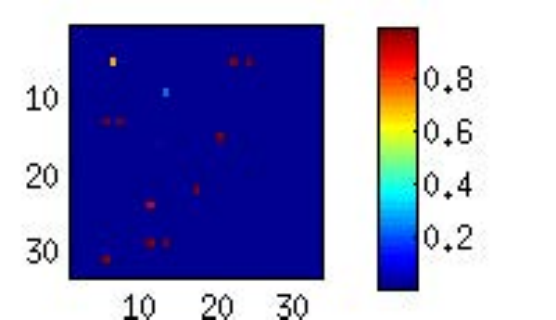}}
      \caption{(mismatched) Non-blind result with a mismatched MRFM type PSF.}
    \label{fig:NONmyopic_x11_MRFM_figImage}
\end{figure}

The results from the variational deconvolution
algorithm with a mismatched Gaussian PSF and a MRFM type PSF are
presented in Fig.~\ref{fig:NONmyopic_x11_Gaussian_figImage} and
Fig.~\ref{fig:NONmyopic_x11_MRFM_figImage}, respectively;
the relevant PSFs and observations are presented in
Fig.~\ref{fig:myopic_x11_Gaussian_figTrueObsPSFs} in Section
\ref{ssec:sim_Gaussian} and in
Fig.~\ref{fig:myopic_x11_MRFM_figTrueObsPSFs} in Section
\ref{ssec:sim_MRFM}, respectively. Compared with the results of our
VB semi-blind algorithm (Algorithm~\ref{algo:VAiter}), shown in
Fig.~\ref{fig:myopic_x11_Gaussian_figImage} and
Fig.~\ref{fig:myopic_x11_MRFM_figImage}, the reconstructed images
from the mismatched non-blind VB algorithm
in Fig.~\ref{fig:NONmyopic_x11_Gaussian_figImage} and
Fig.~\ref{fig:NONmyopic_x11_MRFM_figImage}, respectively,
inaccurately estimate signal locations and blur most of the non-zero
values.

Additional experiments (not shown here)  establish that the PSF estimator is very accurate when the algorithm is initialized with the true image.


\subsection{Comparison with other algorithms}

To quantify the comparison, we performed experiments with
the same set of four sparse images and the MRFM type PSFs as used in \citep{Park2012}. By
generating 100 different noise realizations for 100 independent trials with each
true image, we measured errors according to various criteria. We tested four
sparse images with sparsity levels $\| \bfx\|_0  = 6, 11, 18, 30$.

\begin{figure}[h!]
 \centering
 \subfigure[$\| \hat \bfx \|_0/\|\bfx\|_0$ ]{\label{fig:norm0Plot}
  \includegraphics[trim = 10mm 10mm 0mm 0mm, width=\figwidthsmall]{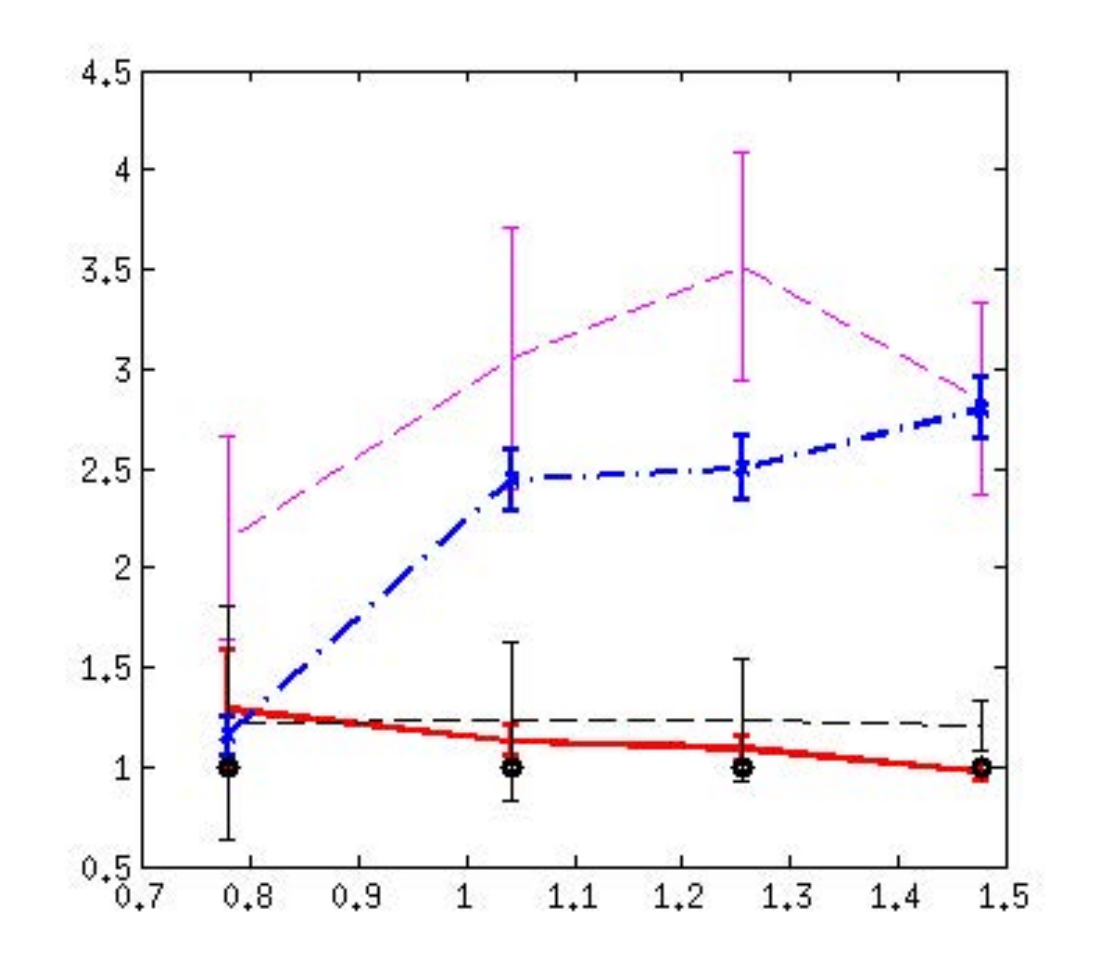}}
 \hspace{0.01\textwidth}
 \subfigure[$\| \frac{\bfx}{\|\bfx\|} - \frac{\hat \bfx}{\|\hat \bfx\|} \|_2^2/\|\bfx\|_0$]{\label{fig:errtPlot}
  \includegraphics[trim = 10mm 10mm 0mm 0mm, width=\figwidthsmall]{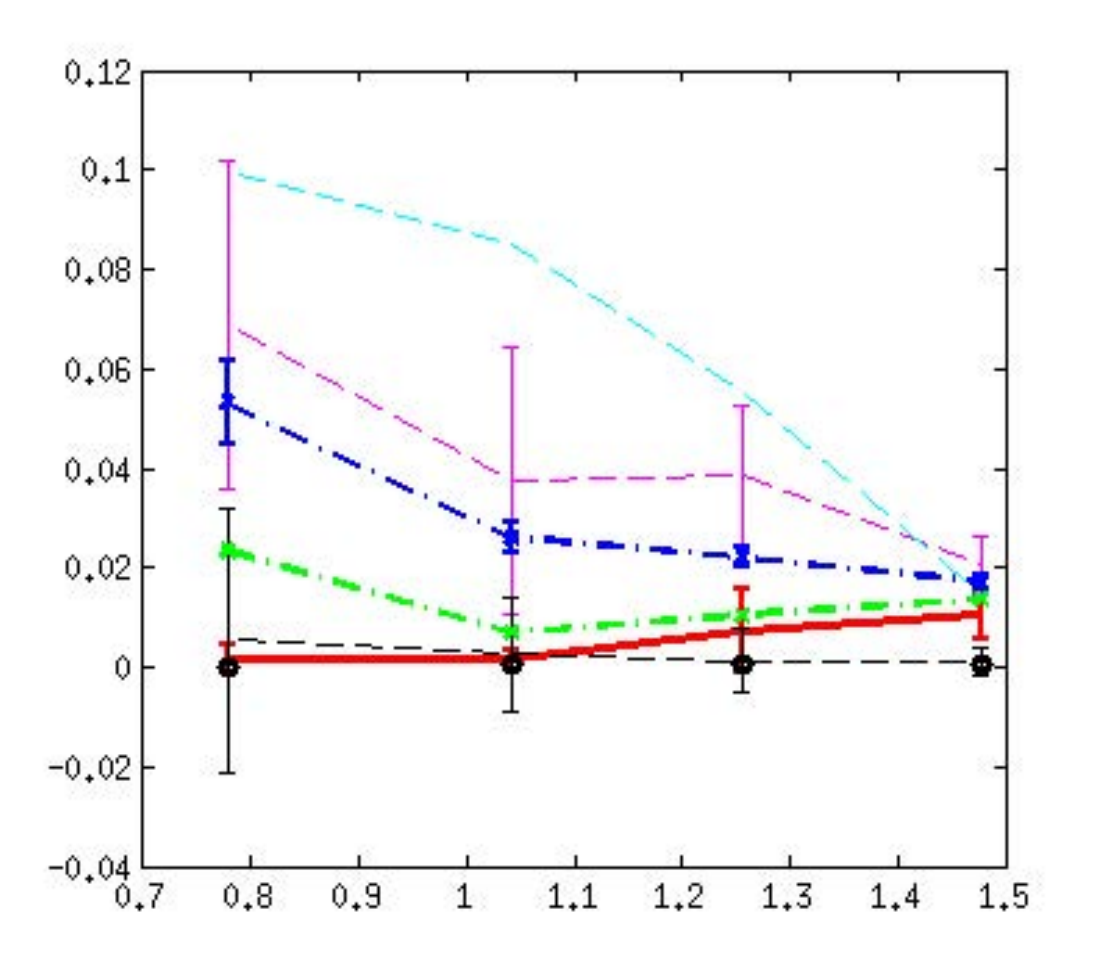}}
 \subfigure[$\| \bfy - \hat \bfy \|_2^2 /\|\bfx\|_0$]{\label{fig:errPlot}
  \includegraphics[trim = 10mm 10mm 0mm 0mm, width=\figwidthsmall]{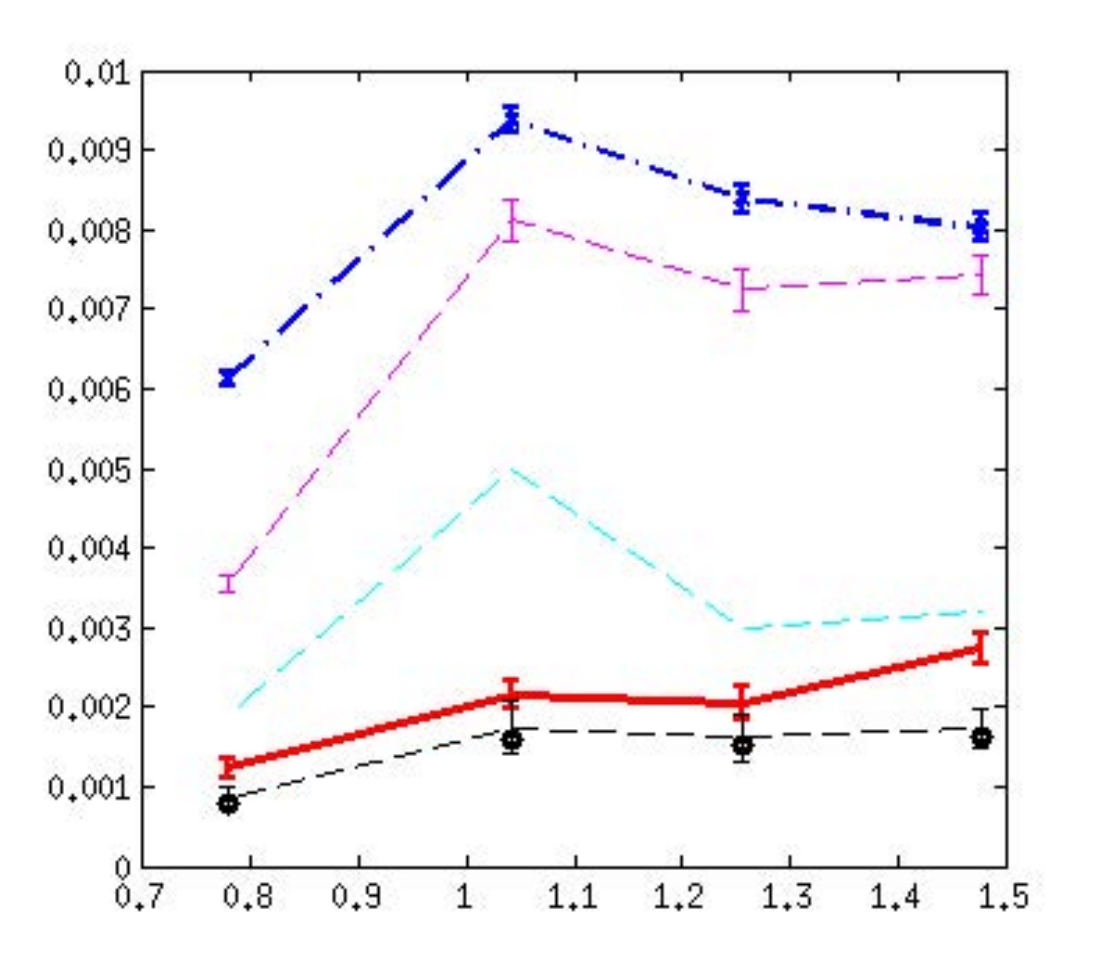}}
 \hspace{0.01\textwidth}
 \subfigure
 [$\| \frac{\hat \psf}{\|\hat \psf\|} - \frac{\psf}{\|\psf\|} \|^2_2$ ]{\label{fig:psferrPlot}
  \includegraphics[trim = 10mm 10mm 0mm 0mm, width=\figwidthsmall]{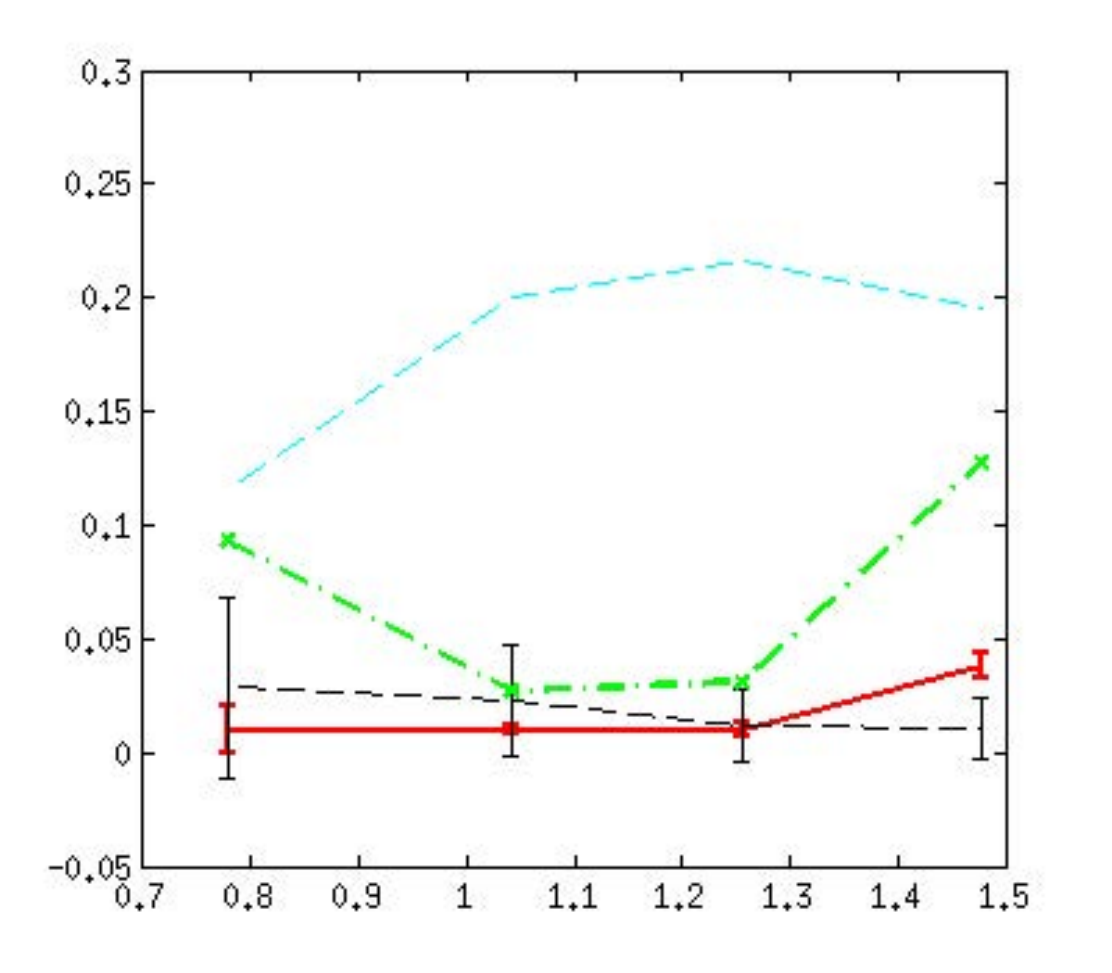}}
 \caption{
For various image sparsity levels (x-axis: $\log_{10} \|\bfx\|_0$), performance of several blind, semi-blind, and nonblind deconvolution algorithms: the proposed method (red), AM (blue), Almeida's method (green), Tzikas's method (cyan), semi-blind MC (black), mismatched nonblind MC (magenta). Errors are illustrated with standard deviations. 
(a): Estimated sparsity. Normalized true level is $1$ (black circles).  
(b): Normalized error in reconstructed image. For the lower bound, information about the true PSF is only available to the oracle IST (black circles). 
(c): Residual (projection) error. The noise level appears in black circles. 
(d): PSF recovery error, as a performance gauge of our semi-blind method. At
 the initial stage of the algorithm, $\| \frac{\psf_0}{\|\psf_0\|} - \frac{\psf}{\|\psf\|} \|^2_2 = 0.5627$. 
(Some of the sparsity measure and residual errors
 are too large to be plotted together with results from other algorithms.)}
 \label{fig:errorbars}
\end{figure}

Under these
criteria\footnote{
Note that the $\ell_0$ norm has been normalized.
The true image has value 1; $\|\hat\bfx\|_0/\|\bfx\|_0$ is used for
MCMC method; $\EE\left[w\right] \times N/\|\bfx\|_0$ for variational method
since this method does not produce zero pixels but $\EE\left[w\right]$. \\
Note also that, for our simulated data, the (normalized) true noise
levels are  $\|\bfn\|^2/\|\bfx \|_0 = 0.1475, 0.2975, 0.2831 ,
0.3062$ for $\|\bfx \|_0= 6, 11, 18, 30$, respectively.},
Fig.~\ref{fig:errorbars} visualizes the reconstruction error performance
for several measures of error. From these figures we conclude that
the VB semi-blind algorithm performs at least as well as the previous MCMC semi-blind algorithm.
In addition, the VB method outperforms AM \citep{Herrity2008} and the mismatched non-blind MCMC \citep{Dobigeon2009a} methods.
 In
terms of PSF estimation, for very sparse images the VB semi-blind method seems to outperform the MCMC method.
Also, the proposed VB semi-blind method
converges more quickly and requires fewer iterations.
For example, the VB semi-blind algorithm converges in approximately 9.6
seconds after 12 iterations, but the previous MCMC algorithm takes
more than 19.2 seconds after 40 iterations to achieve convergence\footnote{
The convergence here is defined as the state where the change in estimation curves over time is negligible.}.

In addition, we made comparisons
between our sparse image reconstruction method and other state-of-the-art blind deconvolution methods
 \citep{Babacan2009,Amizic2010,Almeida2010,Tzikas2009, Fergus2006,Shan2008}, as shown in our previous work \citep{Park2012}. These algorithms were
 initialized with the nominal, mismatched PSF and were applied
 to the same sparse image as our experiment above.
For a fair comparison, we made a sparse prior modification in the image model of other algorithms, as needed. Most of these methods do not assume or fit into the sparse model in our experiments, thus leading to poor performance in terms of image and PSF estimation errors.
Among these tested algorithms, two of them, proposed by Tzikas \emph{et al.} \citep{Tzikas2009} and Almeida \emph{et al.} \citep{Almeida2010}, produced non-trivial and convergent solutions and the \red{ corresponding results} are compared to ours in Fig.~\ref{fig:errorbars}.
By using basis kernels the method proposed by Tzikas \emph{et al.} \citep{Tzikas2009} uses a similar PSF model to ours. Because a sparse image prior is not assumed in their algorithm \citep{Tzikas2009}, we applied their suggested PSF model along with our sparse image prior for a fair comparison.
The method proposed by Almeida \emph{et al.} \citep{Almeida2010} exploits the sharp edge property in natural images and uses initial, high regularization for effective PSF estimation.
Both of these perform worse than our VB method as seen in Fig.~\ref{fig:errorbars}. The remaining algorithms \citep{Babacan2009,Amizic2010,Fergus2006,Shan2008}, which focus on photo image reconstruction or motion blur, either produce a trivial solution ($\hat\bfx \approx \bfy$) or are a special case of Tzikas's model \citep{Tzikas2009}.

To show lower bound our myopic reconstruction
algorithm, we used the Iterative Shrinkage/Thresholding (IST) algorithm with a true PSF. This algorithm effectively restores sparse images with a sparsity constraint \citep{Daubechies2004}.
We demonstrate comparisons of the computation time\footnote{Matlab is used under Windows 7 Enterprise and HP-Z200 (Quad 2.66 GHz)
platform.} of our proposed reconstruction algorithm to that of others in Table \ref{tab:time}.


{
\begin{table}[h!]
\caption{\label{tab:time} Computation time of algorithms (in seconds), for the data in Fig.~\ref{fig:myopic_x11_MRFM_figTrueObsPSFs}. }
\begin{center}
\renewcommand{\arraystretch}{1.2}
\begin{tabular}{|c|c|}
\hline
\hline Our method  & 9.58 \\
\hline semi-blind MC \citep{Park2012}   & 19.20 \\
\hline Bayesian nonblind \citep{Dobigeon2009a} & 3.61  \\
\hline  AM  \citep{Herrity2008} &  0.40\\
\hline   Almeida's method \citep{Almeida2010} &5.63 \\
\hline   Amizic's method \citep{Amizic2010} & 5.69\\
\hline  Tzikas's method \citep{Tzikas2009} &20.31\\
\hline (oracle) IST \citep{Daubechies2004}  & 0.09\\ 
\hline \hline
\end{tabular}
\end{center}
\end{table}
}


\subsection{Application to tobacco mosaic virus (TMV) data}

\begin{figure}[h]
  \centering
  \subfigure[TMV raw data.]{\label{fig:virus_obs}
    \includegraphics[width=0.90\textwidth]{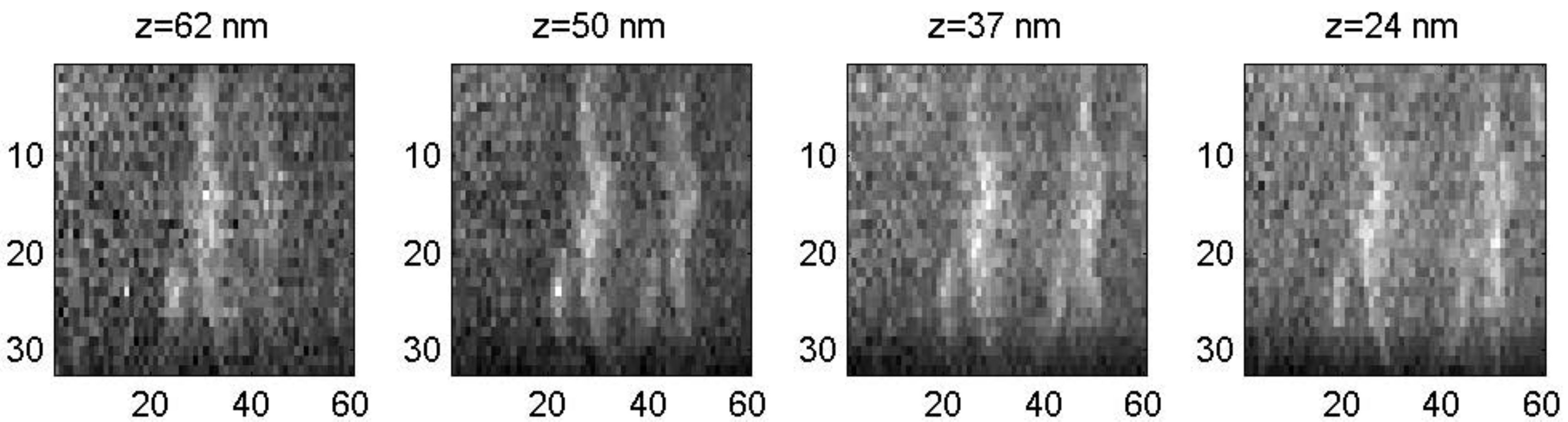}}
  \subfigure[VB estimate]{\label{fig:myopic_virus_image}
    \includegraphics[width=0.17\textwidth,height=0.17\textwidth]{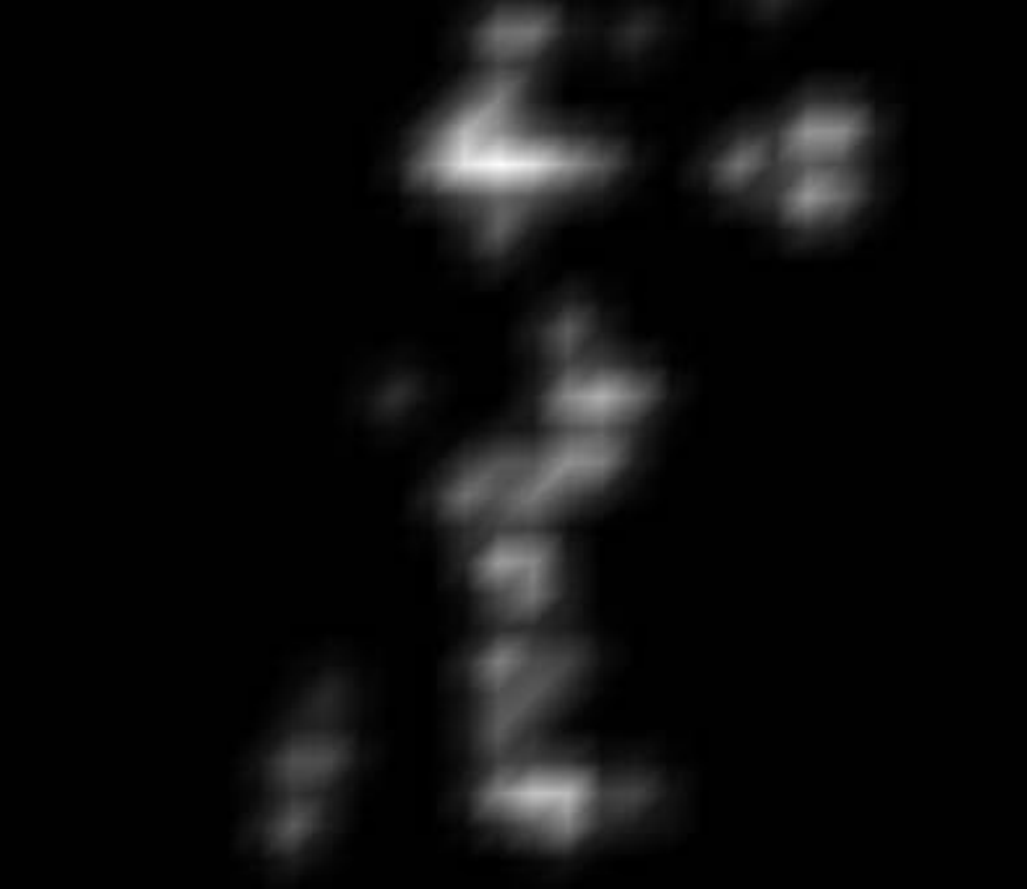}}
  \hspace{0.03\textwidth}
  \subfigure[MC estimate]{\label{fig:myopic_virus_image_MCMC}
    \includegraphics[width=0.17\textwidth,height=0.17\textwidth]{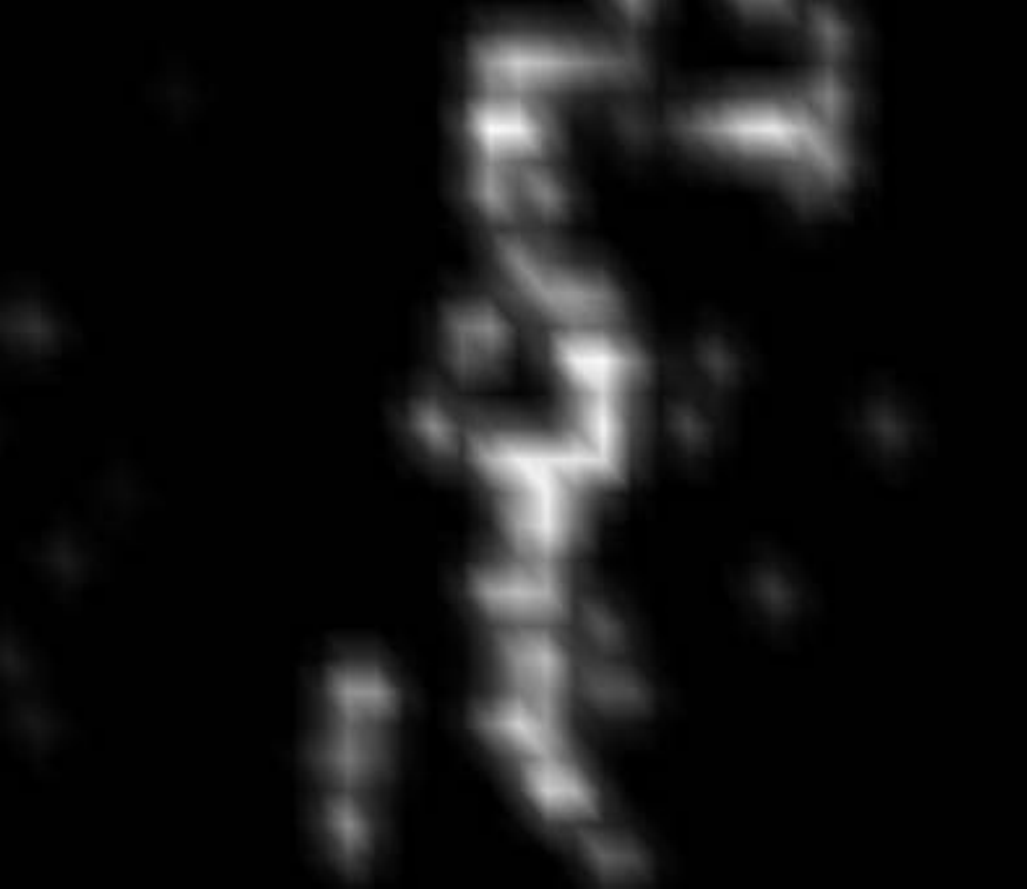}}
  \hspace{0.02\textwidth}
  \subfigure[SEM \citep{Degen2009}]{\label{fig:EMI_virus}
    \includegraphics[width=0.17\textwidth,height=0.17\textwidth]{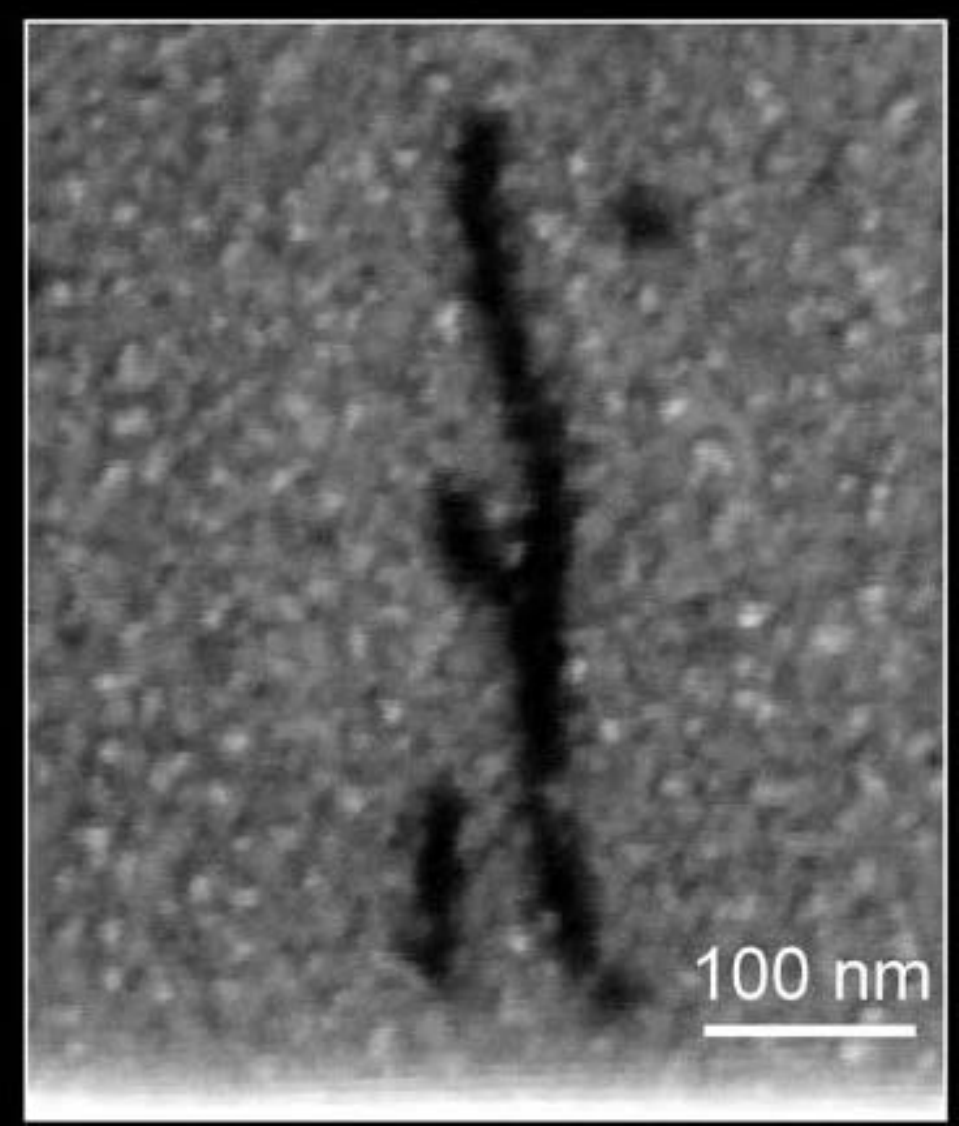}}
  \caption{(a) TMV raw data, (b) estimated virus image by VB, (c) estimated virus image by MCMC \citep{Park2012}, and (d) virus image from electron microscope \citep{Degen2009}. }
  \label{fig:virus}
\end{figure}

We applied the proposed variational semi-blind sparse deconvolution algorithm to the tobacco mosaic virus data, made available by our IBM collaborators \citep{Degen2009}, shown in the first row in Fig.~\ref{fig:virus}. Our algorithm is easily modifiable to these 3D raw image data and 3D PSF with an additional dimension in dealing with basis functions to evaluate each voxel value $x_i$. The noise is assumed Gaussian \citep{Rugar1992,Degen2009} and the four PSF bases are obtained by the procedure proposed in
\ref{ssec:PSFbases} with the physical MRFM PSF model and the nominal parameter values \citep{Mamin2003}.
The reconstruction of the 6th layer is shown in Fig.~\ref{fig:myopic_virus_image},
and is consistent with the results obtained by other methods. (see \citep{Park2012,Dobigeon2009a}.) The estimated deviation in PSF is small, as predicted in \citep{Park2012}. 

\red{
While they now exhibit similar smoothness, the VB and MCMC images are still somewhat different since each algorithm follows different iterative trajectory in the high dimensional space of 3D
images, thus converging possibly to slightly different stopping points near the maximum of the surrogate distribution. 
We conclude that the two images from VB and MCMC are comparable in that both represent the
2D SEM image well, but VB is significantly faster.
}

\subsection{Discussion}
\label{ssec:discussion}

\red{
In blind deconvolution, joint identifiability is a common issue. For example, 
because of scale ambiguity, the unicity cannot be guaranteed in a general setting. 
It is not proven in our solution either. However, the shift/time ambiguity issue noticed in \citep{Ge2011} is implicitly addressed in our method using a nominal and basis PSFs. Moreover, our constraint on the PSF space using a basis approach effectively excludes a delta function as a PSF solution, thus avoiding the trivial solution. Secondly, the PSF solution is restricted
to this linear spanning space, starting form the initial, mismatched PSF. 
We can, therefore, reasonably expect that the solution provided
by the algorithm is close to the true PSF, away from the trivial solution or the initial PSF.
\\
To resolve scale ambiguity in a MCMC Bayesian framework, stochastic samplers are proposed   in \citep{Ge2011} by imposing a fixed variance on a certain distribution\footnote{We note that this MCMC method designed for 1D signal deconvolution
 is not efficient for analyzing 2D and 3D images, since the grouped and marginalized samplers are usually slow to converge requiring hundreds of iterations \citep{Ge2011}.  
}. 
Another approach to resolve the scale ambiguity is to assume a hidden scale variable that is multiplied to the PSF and dividing the image (or vice versa.), where the scale is drawn along each iteration of the Gibbs sampler \citep{Vincent2010}. 
}

\section{Conclusion}
\label{sec:concl}
We suggested a novel variational solution to a semi-blind sparse deconvolution problem.
Our method uses Bayesian inference for image and PSF restoration with a sparsity-inducing image prior via the variational Bayes approximation. Its power in automatically producing all required parameter values from the data merits further attention for the extraction of image properties and retrieval of necessary features.

From the simulation results, we conclude that the performance of the VB method competes with MCMC methods in sparse image estimation, while requiring fewer computations.
Compared to a non-blind algorithm whose mismatched PSF leads to imprecise and blurred signal locations in the restored image, the VB semi-blind algorithm correctly produces sparse image estimates.
The benefits of this solution compared to the previous solution \citep{Park2012} are faster convergence and stability of the method.


\renewcommand{\baselinestretch}{1.1}
\small

\appendix

\section{Useful Distributions}

\subsection{Inverse Gamma Distribution}
\label{appen:IG}
The density of an inverse Gamma random variable $X \sim \calI\calG(a,b) $ is
$\dfrac{b^a}{\Gamma(a)} x^{-a-1} \exp(-\frac{b}{x})$, for $x \in (0,\infty)$. $\EE X^{-1} = a/b$ and $\EE \ln(X) = \ln(b) - \psi(a)$.

\subsection{Beta Distribution}
\label{appen:Beta} The density of a Beta random variable $X \sim
\calB(a,b)$ is $\dfrac{\Gamma(a)\Gamma(b)}{\Gamma(a+b)} x^{b-1}
(1-x)^{a-1}$, for $x \in (0,1)$, with $\Gamma(c) = \int_0^\infty
t^{c-1} e^{-t} dt$. The mean of $\calB(a,b)$ is $\frac{b}{a+b}$ and
$\EE \ln( \calB(a,b) ) = \psi(b) - \psi(a+b)$, where $\psi$ is a
digamma function.

\subsection{Positively Truncated Gaussian Distribution}
\label{appen:TGaussian}
The density of a truncated Gaussian random variable $x_i$ is denoted by $x_i \sim \calN_+(x_i;\mu, \eta)$,
and its statistics used in the paper are
\begin{align*}
\EE\left[x_i| x_i>0\right] & = \EE\left[\calN_+(x_i;\mu, \eta)\right] \\
                 & = \mu + \sqrt{\eta} \frac{\phi(-\mu/\sqrt{\eta})}{1-\Phi_0(-\mu/\sqrt{\eta})}, \\
\EE \left[x_i^2 | x_i>0 \right] & = \mathrm{var}[x_i | x_i>0] + (\EE \left[x_i | x_i>0 \right])^2 \\
& = \eta + \mu (\EE\left[x_i| x_i>0\right]) ,
\end{align*}
 where $\Phi_0$ is a cumulative distribution function for the standard normal distribution.

\section{Derivations of $q(\cdot)$}
\label{appen:deriv_q}

In this section, we derive the posterior densities defined by variational Bayes framework in Section \ref{sec:VA}.

\subsection{Derivation of $q(\bfc)$}

We denote the expected value of the squared residual term by $R= \EE \| \bfy - \bfH\bfx\|^2$.
For $c_l, l=1,\ldots,K$,
\begin{align*}
R = & \EE \| \bfy - \bfH^0 \bfx - \sum_{l \neq j} \bfH^l \bfx c_l - \bfH^j \bfx c_j\|^2 \\
  = & c_j^2 \langle \bfx^T {\bfH^j}^T \bfH^j \bfx \rangle - 2 c_j  \langle \bfx^T {\bfH^j}^T \bfy - \bfx {\bfH^j}^T \bfH^0 \bfx  \\
&   - \sum_{l \neq j} \bfx^T {\bfH^j}^T \bfH^l c_l \bfx \rangle + \mathrm{const},
\end{align*}
where $\bfH^j$ is the convolution matrix corresponding to the
convolution with $\psf_j$. For $i\neq j$ and $i,j > 0$, $\EE
(\bfH^i \bfx)^T (\bfH^j \bfx) = tr({\bfH^i}^T \bfH^j ( \mathrm{cov}(\bfx) +
\langle\bfx\rangle \langle\bfx^T \rangle)) = (\bfH^i
\langle\bfx\rangle)^T  (\bfH^j\langle\bfx\rangle)$, since $tr
({\bfH^i}^T \bfH^j \mathrm{cov}(\bfx)) = tr( {\bfH^i D}^T \bfH^j D) =
\sum_k d_k^2 \bfh_k^i \bfh_k^j = 0$.
Here, $\mathrm{cov}(\bfx)$ is approximated as a diagonal matrix $D^2 =
\mathrm{diag}(d_1^2 ,\ldots, d_n^2)$.
This is reasonable, especially when the expected recovered
signal $\hat\bfx$ exhibits high sparsity.
Likewise, $\EE (\bfH^0 \bfx)^T (\bfH^j \bfx) =
\psf_0^T \psf_j \sum_i \mathrm{var}[x_i] + (\bfH^0
\langle\bfx\rangle)^T (\bfH^j\langle\bfx\rangle)$ and $\EE (\bfH^j
\bfx)^T (\bfH^j \bfx) = \| \psf_j \|^2 \sum_i \mathrm{var}[x_i] + \|
\bfH^j \langle\bfx\rangle \|^2 $.

Then, we factorize $\EE \left[- \frac{R}{2\sigma^2} \right] = -\frac{ (c_j -
\mu_{c_j} )^2 }{2 \sigma_{c_j} } $, with $\mu_{c_j} = \frac{ \langle
\bfx^T {\bfH^j}^T \bfy - \bfx {\bfH^j}^T \bfH^0 \bfx - \sum_{l \neq
j} \bfx^T {\bfH^j}^T \bfH^l c_l \bfx \rangle}{\langle \bfx^T
{\bfH^j}^T \bfH^j \bfx \rangle}$, $1/\sigma_{c_j} = \langle
1/\sigma^2 \rangle \langle \bfx^T {\bfH^j}^T \bfH^j \bfx \rangle$.

If we set the prior, $p(c_j)$, to be a uniform distribution over a wide range of the real line that covers error tolerances, we obtain a normally distributed variational density $q(c_j) = \phi(\mu_{c_j},\sigma_{c_j})$ with its mean $\mu_{c_j}$ and variance $\sigma_{c_j}$ defined above, because $\ln q(c_j) =  \EE \left[- \frac{R}{2\sigma^2}\right] $.
By the independence assumption, $q(\bfc) = \prod q(c_j)$, so $q(\bfc)$ can be easily evaluated.

\subsection{Derivation of $q(\sigma^2)$}

We evaluate $R$ ignoring edge effects; $R = \|\bfy -
\langle\bfH\rangle \langle\bfx\rangle \|^2 + \sum \mathrm{var}[x_i]
[ \| \langle \psf \rangle \|^2  + \sum_l \sigma_{c_l} \|\psf_l\|^2]
+ \sum_l \sigma_{c_l}  \|{\bfH^l\langle\bfx\rangle}\|^2$.
$\|\psf\|^2$ is a kernel energy in $\ell_2$ sense and the variance
terms add uncertainty, due to the uncertainty in $\psf$, to the
estimation of density.
Applying \eqref{eq:KL},
(ignoring constants)
\begin{align*}
\ln q(\sigma^2)   & = \EE_{\backslash \sigma^2} \left[\ln p(\bfy|\bfx ,\bfc, \sigma^2) p(\sigma^2) p(\bfx|a, w) p(w)p(a) \right]  \\
                   & = \EE_{\bfx,\bfc} \left[\ln  p(\bfy|\bfx , \sigma^2 )\right] + \ln p(\sigma^2)  \\
                   & =  - \frac{\EE_{\bfx,\bfc} \left[\|\bfy - \bfH\bfx\|^2\right]}{2\sigma^2} -\frac{P}{2} \ln\sigma^2 + \ln p(\sigma^2). \\
 \calI\calG(\tilde\varsigma_0,\tilde\varsigma_1) & \triangleq q(\sigma^2) = \calI\calG(P/2 + \varsigma_0, \langle \|\bfy-\bfH\bfx\|^2 \rangle /2 + \varsigma_1) .
\end{align*}
($\EE_{\backslash \sigma^2}$ denotes expectation with respect to all variables except $\sigma^2$.)

\subsection{Derivation of $q(\bfx)$}\label{apssec:q_x}

For $x_i, i=1,\ldots,N$, $ R = \EE \| \bfe_i - \bfh_i x_i \|^2$ with $\bfe_i = \bfy - \bfH \bfx_{-i} = \bfy -\bfH^0 \bfx_{-i} - \sum_l \bfH^l c_l \bfx_{-i}$, $\bfh_i = [\bfH^0 + \sum\bfH^l c_l]_i = \bfh^0_i + \sum \bfh^l_i c_l =$ ($i$th column of $\bfH$). Ignoring constants, $R = \langle \|\bfh_i\|^2 \rangle x_i^2 -2 \langle \bfh_i^T \bfe_i \rangle x_i$.

Using the orthogonality of the kernel bases and uncorrelatedness of
$c_l$'s, we derive the following terms (necessary to evaluate $R$):
$\langle \|\bfh_i\|^2 \rangle = \| \bfh^0_i \|^2 + \sum_{l}
\sigma_{c_l} \| {\bfh^l}_i \|^2 $ and, $\langle \bfh_i^T \bfe_i
\rangle = \langle \bfh_i^T \rangle (\bfy - \langle\bfH\rangle
\langle\bfx_{-i}\rangle) - \sum_l \mathrm{var}[c_l] {\bfh^l_i}^T
\bfH^l \langle\bfx_{-i}\rangle $.

Then, $\mathrm{var}[x_i] = w_i' \EE \left[x_i^2 | x_i>0\right] - w_i'^2
(\EE\left[x_i| x_i>0\right])^2  $, $\EE\left[x_i\right] = w_i' \EE\left[x_i|x_i > 0\right]$, where
$w_i' = q(z_i = 1)$ is the posterior weight for the normal
distribution and $1-w_i'$ is the weight for the delta function. The required
statistics of $x_i$ that are used to derive the distribution above
are obtained by applying  \ref{appen:TGaussian}.

\subsection{Derivation of $q(\bfz)$}

To derive $q(z_i=1) = \langle z_i \rangle$, we evaluate the unnormalized version $\hat q(z_i)$ of $q(z_i)$ and normalize it.
$\ln \hat q(z_i=1) = \EE_{\backslash z_i} \left[ -\frac{\|\bfe_i - \bfh_i x_i \|^2}{2 \sigma^2} - \ln a - \frac{x_i}{a} + \ln w  \right] $ with ${x_i \sim N_+(\mu_i,\eta_i)}$ and $\ln \hat q(z_i=0) = \EE_{\backslash z_i} \left[ -\frac{\|\bfe_i  \|^2}{2 \sigma^2} + \ln (1-w)  \right] $ with ${ x_i = 0}$. The normalized version of the weight is $q(z_i=1) = {1}/[{1+ C^\prime_i }]$.
$C^\prime_i = \exp(\ln \hat q(z_i=0)-\ln \hat q(z_i=1)) = \exp( C_i/2 \times \langle 1/\sigma^2 \rangle + \mu \langle 1/a \rangle + \langle \ln a \rangle + \langle \ln(1-w)  - \ln w \rangle = \exp(C_i/2 \times \tilde\varsigma_0/\tilde\varsigma_1 + \mu \tilde\alpha_0/\tilde\alpha_1 + \ln \tilde\alpha_1 - \psi(\tilde\alpha_0) + \psi(\tilde\beta_0) - \psi(\tilde\beta_1) )$. $\psi$ is a digamma function and $C_i= \langle \|\bfh_i\|^2 \rangle (\mu_i^2 + \eta_i) -2 \langle\bfe_i^T\bfh_i \rangle \mu_i$.

\renewcommand{\baselinestretch}{0.6}
\scriptsize
\bibliographystyle{elsarticle-num}
\bibliography{biblio_rev}

\end{document}

%% file: notations.tex
\input com_notations.tex

\newcommand{\MATobs}{\bfY}
\newcommand{\Vobs}{\mathbf{y}}
\newcommand{\obs}[1]{y_{#1}}

\newcommand{\MATima}{\bfX}
\newcommand{\Vima}{\bfx}
\newcommand{\ima}[1]{x_{#1}}

\newcommand{\dimm}{M}
\newcommand{\dimn}{P}
\newcommand{\dimima}{n}

\newcommand{\ftrans}[2]{T\left(#1,#2\right)}
\newcommand{\MATtrans}{\mathbf{H}}
\newcommand{\Vtrans}[1]{\mathbf{h}_{i}}

\newcommand{\psf}{\boldsymbol{\kappa}}
\newcommand{\Vpsfparam}{\boldsymbol{\lambda}}
\newcommand{\psfparam}[1]{\lambda_{#1}}

\newcommand{\MATnoise}{\mathbf{N}}
\newcommand{\Vnoise}{\mathbf{n}}
\newcommand{\noisevar}{{\sigma^2}}

\newcommand{\hypervect}{\boldsymbol{\Phi}}
\newcommand{\paramvect}{\boldsymbol{\theta}}

\newcommand{\Valpha}{\boldsymbol{\alpha}}
\newcommand{\Vomega}{\boldsymbol{\omega}}

\newcommand{\sample}[2]{#1^{(#2)}}
\newcommand{\samplebis}[2]{{#1}^{(#2)}}
\newcommand{\samplenoisevar}[1]{{\widetilde{\sigma}}^{2(#1)}}

\newcommand{\norm}[1]{\left\|#1\right\|}

\newcommand{\R}{\mathds{R}}

\newcommand{\dirac}[1]{\delta\left({#1}\right)}

\newcommand{\inv}{^{-1}}

\newcommand{\herm}{^{H}}

\newcommand{\transp}{^T}

\newcommand{\etr}{\mathrm{etr}}

\newcommand{\Ndistr}[1]{\mathcal{N}\left(#1\right)}

\newcommand{\Vun}{{\boldsymbol{1}}}
\newcommand{\Vzero}{\boldsymbol{0}}
\newcommand{\Id}[1]{\textbf{I}_{#1}}
\newcommand{\Indicfun}[2]{\textbf{1}_{#1}\left(#2\right)}

\newenvironment{algogo}[1]{
\smallskip
\noindent \hrule\vspace{0.2\baselineskip} \hrule
\smallskip
\begin{small}
\refstepcounter{algo} \center{\bf \textsc{Algorithm \thealgo:}}
\\{\center{\bf #1}}
\smallskip
\flushleft
 } {
\end{small}
\bigskip
\hrule\vspace{0.2\baselineskip} \hrule
\smallskip }

\newcounter{algo}
\renewcommand{\thealgo}{\arabic{algo}}

%% file: com_notations.tex

\def\bfa{{\mathbf{a}}}
\def\bfb{{\mathbf{b}}}
\def\bfc{{\mathbf{c}}}
\def\bfd{{\mathbf{d}}}
\def\bfe{{\mathbf{e}}}
\def\bff{{\mathbf{f}}}
\def\bfg{{\mathbf{g}}}
\def\bfh{{\mathbf{h}}}
\def\bfi{{\mathbf{i}}}
\def\bfj{{\mathbf{j}}}
\def\bfk{{\mathbf{k}}}
\def\bfl{{\mathbf{l}}}
\def\bfm{{\mathbf{m}}}
\def\bfn{{\mathbf{n}}}
\def\bfo{{\mathbf{o}}}
\def\bfp{{\mathbf{p}}}
\def\bfq{{\mathbf{q}}}
\def\bfr{{\mathbf{r}}}
\def\bfs{{\mathbf{s}}}
\def\bft{{\mathbf{t}}}
\def\bfu{{\mathbf{u}}}
\def\bfv{{\mathbf{v}}}
\def\bfw{{\mathbf{w}}}
\def\bfx{{\mathbf{x}}}
\def\bfy{{\mathbf{y}}}
\def\bfz{{\mathbf{z}}}

\def\bfA{{\mathbf{A}}}
\def\bfB{{\mathbf{B}}}
\def\bfC{{\mathbf{C}}}
\def\bfD{{\mathbf{D}}}
\def\bfE{{\mathbf{E}}}
\def\bfF{{\mathbf{F}}}
\def\bfG{{\mathbf{G}}}
\def\bfH{{\mathbf{H}}}
\def\bfI{{\mathbf{I}}}
\def\bfJ{{\mathbf{J}}}
\def\bfK{{\mathbf{K}}}
\def\bfL{{\mathbf{L}}}
\def\bfM{{\mathbf{M}}}
\def\bfN{{\mathbf{N}}}
\def\bfO{{\mathbf{O}}}
\def\bfP{{\mathbf{P}}}
\def\bfQ{{\mathbf{Q}}}
\def\bfR{{\mathbf{R}}}
\def\bfS{{\mathbf{S}}}
\def\bfT{{\mathbf{T}}}
\def\bfU{{\mathbf{U}}}
\def\bfV{{\mathbf{V}}}
\def\bfW{{\mathbf{W}}}
\def\bfX{{\mathbf{X}}}
\def\bfY{{\mathbf{Y}}}
\def\bfZ{{\mathbf{Z}}}


\def\bbA{{\mathbb{A}}}
\def\bbB{{\mathbb{B}}}
\def\bbC{{\mathbb{C}}}
\def\bbD{{\mathbb{D}}}
\def\bbE{{\mathbb{E}}}
\def\bbF{{\mathbb{F}}}
\def\bbG{{\mathbb{G}}}
\def\bbH{{\mathbb{H}}}
\def\bbI{{\mathbb{I}}}
\def\bbJ{{\mathbb{J}}}
\def\bbK{{\mathbb{K}}}
\def\bbL{{\mathbb{L}}}
\def\bbM{{\mathbb{M}}}
\def\bbN{{\mathbb{N}}}
\def\bbO{{\mathbb{O}}}
\def\bbP{{\mathbb{P}}}
\def\bbQ{{\mathbb{Q}}}
\def\bbR{{\mathbb{R}}}
\def\bbS{{\mathbb{S}}}
\def\bbT{{\mathbb{T}}}
\def\bbU{{\mathbb{U}}}
\def\bbV{{\mathbb{V}}}
\def\bbW{{\mathbb{W}}}
\def\bbX{{\mathbb{X}}}
\def\bbY{{\mathbb{Y}}}
\def\bbZ{{\mathbb{Z}}}


\def\dsA{{\mathds{A}}}
\def\dsB{{\mathds{B}}}
\def\dsC{{\mathds{C}}}
\def\dsD{{\mathds{D}}}
\def\dsE{{\mathds{E}}}
\def\dsF{{\mathds{F}}}
\def\dsG{{\mathds{G}}}
\def\dsH{{\mathds{H}}}
\def\dsI{{\mathds{I}}}
\def\dsJ{{\mathds{J}}}
\def\dsK{{\mathds{K}}}
\def\dsL{{\mathds{L}}}
\def\dsM{{\mathds{M}}}
\def\dsN{{\mathds{N}}}
\def\dsO{{\mathds{O}}}
\def\dsP{{\mathds{P}}}
\def\dsQ{{\mathds{Q}}}
\def\dsR{{\mathds{R}}}
\def\dsS{{\mathds{S}}}
\def\dsT{{\mathds{T}}}
\def\dsU{{\mathds{U}}}
\def\dsV{{\mathds{V}}}
\def\dsW{{\mathds{W}}}
\def\dsX{{\mathds{X}}}
\def\dsY{{\mathds{Y}}}
\def\dsZ{{\mathds{Z}}}

\def\calh{{\mathcal{h}}}
\def\calU{{\mathcal{U}}}
\def\calu{{\mathcal{u}}}
\def\calS{{\mathcal{S}}}
\def\calV{{\mathcal{V}}}
\def\calv{{\mathcal{v}}}
\def\calP{{\mathcal{P}}}
\def\calA{{\mathcal{A}}}
\def\calB{{\mathcal{B}}}
\def\calC{{\mathcal{C}}}
\def\calD{{\mathcal{D}}}
\def\calE{{\mathcal{E}}}
\def\calF{{\mathcal{F}}}
\def\calG{{\mathcal{G}}}
\def\calH{{\mathcal{H}}}
\def\calI{{\mathcal{I}}}
\def\calJ{{\mathcal{J}}}
\def\calK{{\mathcal{K}}}
\def\calL{{\mathcal{L}}}
\def\calM{{\mathcal{M}}}
\def\calN{{\mathcal{N}}}
\def\calO{{\mathcal{O}}}
\def\calP{{\mathcal{P}}}
\def\calQ{{\mathcal{Q}}}
\def\calR{{\mathcal{R}}}
\def\calS{{\mathcal{S}}}
\def\calT{{\mathcal{T}}}
\def\calU{{\mathcal{U}}}
\def\calV{{\mathcal{V}}}
\def\calW{{\mathcal{W}}}
\def\calx{{\mathcal{x}}}
\def\calX{{\mathcal{X}}}
\def\caly{{\mathcal{y}}}
\def\calY{{\mathcal{Y}}}
\def\calZ{{\mathcal{Z}}}